\documentclass[twocolumn]{aastex701}

\usepackage{natbib}
\usepackage{subcaption}
\usepackage{xspace}
\usepackage{makecell}
\usepackage{amsmath}
\usepackage{hyperref}

\begin{document}

\title{ODIN: Characterizing the Three-dimensional Structure of Two Protocluster Complexes at $z = 3.1$}

\author[0000-0002-9176-7252]{Vandana Ramakrishnan}
\affiliation{Department of Physics and Astronomy, Purdue University, 525 Northwestern Avenue, West Lafayette, IN 47907, USA}
\email[show]{ramakr18@purdue.edu}

\author[0009-0008-3184-304X]{Ashley Ortiz}
\affiliation{Department of Physics and Astronomy, Purdue University, 525 Northwestern Avenue, West Lafayette, IN 47907, USA}
\email{ortiz140@purdue.edu}

\author[0009-0008-4022-3870]{Byeongha Moon}
\affiliation{Korea Astronomy and Space Science Institute, 776 Daedeokdae-ro, Yuseong-gu, Daejeon 34055, Republic of Korea}
\email{bhmoon@kasi.re.kr}

\author[0009-0007-9489-8278]{Eunsoo Jun}
\affiliation{Korea Astronomy and Space Science Institute, 776 Daedeokdae-ro, Yuseong-gu, Daejeon 34055, Republic of Korea}
\email{esjun@kasi.re.kr}

\author{David Schlegel}
\email{djschlegel@lbl.gov}
\affiliation{Lawrence Berkeley National Laboratory, 1 Cyclotron Road, Berkeley, CA 94720, USA}

\author[0000-0003-3004-9596]{Kyoung-Soo Lee}
\affiliation{Department of Physics and Astronomy, Purdue University, 525 Northwestern Avenue, West Lafayette, IN 47907, USA}
\email{soolee@purdue.edu}

\author{Jessica Nicole Aguilar}
\email{jaguilar@lbl.gov}
\affiliation{Lawrence Berkeley National Laboratory, 1 Cyclotron Road, Berkeley, CA 94720, USA}

\author[0000-0003-0570-785X]{Maria Celeste Artale}
\email{mcartale@gmail.com}
\affiliation{Universidad Andres Bello, Facultad de Ciencias Exactas, Departamento de Fisica y Astronomia, Instituto de Astrofisica, Fernandez Concha 700, Las Condes, Santiago RM, Chile}

\author{David Brooks}
\email{david.brooks@ucl.ac.uk}
\affiliation{Department of Physics \& Astronomy, University College London, Gower Street, London, WC1E 6BT, UK}

\author[0009-0000-9347-1933]{Maria Candela Cerdosino}
\email{candelacerdosino@mi.unc.edu.ar}
\affiliation{Instituto de Astronomía Teórica y Experimental (IATE), CONICET-UNC, Laprida 854, X500BGR, Córdoba, Argentina}
\affiliation{Facultad de Matemática, Astronomía, Física y Computación, Universidad Nacional de Córdoba (FaMAF–UNC), Bvd. Medina Allende s/n, Ciudad Universitaria, X5000HUA, Córdoba, Argentina}

\author[0000-0002-1328-0211]{Robin Ciardullo}
\email{rbc3@psu.edu}
\affiliation{Department of Astronomy \& Astrophysics, The Pennsylvania State University, University Park, PA 16802, USA}

\author{Todd Claybaugh}
\email{tmclaybaugh@lbl.gov}
\affiliation{Lawrence Berkeley National Laboratory, 1 Cyclotron Road, Berkeley, CA 94720, USA}

\author[0000-0002-2169-0595]{Andrei Cuceu}
\email{acuceu@lbl.gov}
\affiliation{Lawrence Berkeley National Laboratory, 1 Cyclotron Road, Berkeley, CA 94720, USA}

\author[0000-0002-1769-1640]{Axel de la Macorra}
\email{macorra@fisica.unam.mx}
\affiliation{Instituto de F\'{\i}sica, Universidad Nacional Aut\'{o}noma de M\'{e}xico,  Circuito de la Investigaci\'{o}n Cient\'{\i}fica, Ciudad Universitaria, Cd. de M\'{e}xico  C.~P.~04510,  M\'{e}xico}

\author[0000-0002-4928-4003]{Arjun Dey}
\email{arjun.dey@noirlab.edu}
\affiliation{NSF NOIRLab, 950 N. Cherry Ave., Tucson, AZ 85719, USA}

\author[0000-0002-9811-2443]{Nicole M. Firestone}
\email{nmf82@physics.rutgers.edu}
\affiliation{Department of Physics and Astronomy, Rutgers, the State University of New Jersey, Piscataway, NJ 08854, USA}

\author[0000-0002-3033-7312]{Andreu Font-Ribera}
\email{afont@ifae.es}
\affiliation{Institut de F\'{i}sica d’Altes Energies (IFAE), The Barcelona Institute of Science and Technology, Edifici Cn, Campus UAB, 08193, Bellaterra (Barcelona), Spain}

\author[0000-0002-2890-3725]{Jaime E. Forero-Romero}
\email{je.forero@uniandes.edu.co}
\affiliation{Departamento de F\'isica, Universidad de los Andes, Cra. 1 No. 18A-10, Edificio Ip, CP 111711, Bogot\'a, Colombia}
\affiliation{Observatorio Astron\'omico, Universidad de los Andes, Cra. 1 No. 18A-10, Edificio H, CP 111711 Bogot\'a, Colombia}

\author[0000-0003-1530-8713]{Eric Gawiser}
\email{gawiser@physics.rutgers.edu}
\affiliation{Department of Physics and Astronomy, Rutgers, the State University of New Jersey, Piscataway, NJ 08854, USA}

\author[0000-0001-9632-0815]{Enrique Gaztañaga}
\email{gaztanaga@gmail.com}
\affiliation{Institute of Space Sciences, ICE-CSIC, Campus UAB, Carrer de Can Magrans s/n, 08913 Bellaterra, Barcelona, Spain}
\affiliation{Institute of Cosmology and Gravitation, University of Portsmouth, Dennis Sciama Building, Portsmouth, PO1 3FX, UK}
\affiliation{Institut d'Estudis Espacials de Catalunya (IEEC), c/ Esteve Terradas 1, Edifici RDIT, Campus PMT-UPC, 08860 Castelldefels, Spain}

\author[0000-0001-6842-2371]{Caryl Gronwall}
\email{caryl@astro.psu.edu}
\affiliation{Institute for Gravitation and the Cosmos, The Pennsylvania State University, University Park, PA 16802, USA}
\affiliation{Department of Astronomy \& Astrophysics, The Pennsylvania State University, University Park, PA 16802, USA}

\author[0000-0002-4902-0075]{Lucia Guaita}
\email{lucia.guaita@gmail.com}
\affiliation{Universidad Andres Bello, Facultad de Ciencias Exactas, Departamento de Fisica y Astronomia, Instituto de Astrofisica, Fernandez Concha 700, Las Condes, Santiago RM, Chile}

\author{Gaston Gutierrez}
\email{gaston@fnal.gov}
\affiliation{Fermi National Accelerator Laboratory, PO Box 500, Batavia, IL 60510, USA}

\author[0000-0001-9991-8222]{Sungryong Hong}
\email{shongscience@kasi.re.kr}
\affiliation{Korea Astronomy and Space Science Institute, 776 Daedeokdae-ro, Yuseong-gu, Daejeon 34055, Republic of Korea}

\author[0000-0003-3428-7612]{Ho Seong Hwang}
\email{galaxy79@snu.ac.kr}
\affiliation{SNU Astronomy Research Center, Seoul National University, 1 Gwanak-ro, Gwanak-gu, Seoul 08826, Republic of Korea}
\affiliation{Department of Physics and Astronomy, Seoul National University, 1 Gwanak-ro, Gwanak-gu, Seoul 08826, Republic of Korea}

\author[0009-0003-9748-4194]{Sang Hyeok Im}
\email{sanghyeok.im97@gmail.com}
\affiliation{Department of Physics and Astronomy, Seoul National University, 1 Gwanak-ro, Gwanak-gu, Seoul 08826, Republic of Korea}
\affiliation{Korea Institute for Advanced Study, 85 Hoegi-ro, Dongdaemun-gu, Seoul 02455, Republic of Korea}

\author[0000-0001-6162-3023]{Paulina Troncoso Iribarren}
\email{p.troncoso.iribarren@gmail.com}
\affiliation{Facultad de Ingenieria y Arquitectura, Universidad Central de Chile, Avenida Francisco de Aguirre 0405, 171-0614 La Serena, Coquimbo, Chile}

\author[0000-0002-2770-808X]{Woong-Seob Jeong}
\email{jeongws@kasi.re.kr}
\affiliation{Korea Astronomy and Space Science Institute, 776 Daedeokdae-ro, Yuseong-gu, Daejeon 34055, Republic of Korea}

\author[0000-0003-0201-5241]{Dick Joyce}
\email{richard.joyce@noirlab.edu}
\affiliation{NSF NOIRLab, 950 N. Cherry Ave., Tucson, AZ 85719, USA}

\author[0000-0001-6270-3527]{Ankit Kumar}
\email{ankit4physics@gmail.com}
\affiliation{Departamento de Ciencias Fisicas, Universidad Andres Bello, Fernandez Concha 700, Las Condes, Santiago, Chile}

\author[0000-0002-6731-9329]{Claire Lamman}
\email{lamman.1@osu.edu}
\affiliation{The Ohio State University, Columbus, 43210 OH, USA}

\author[0000-0003-1838-8528]{Martin Landriau}
\email{mlandriau@lbl.gov}
\affiliation{Lawrence Berkeley National Laboratory, 1 Cyclotron Road, Berkeley, CA 94720, USA}

\author[:0000-0001-5342-8906]{Seong-Kook Lee}
\email{s.joshualee@gmail.com}
\affiliation{SNU Astronomy Research Center, Seoul National University, 1 Gwanak-ro, Gwanak-gu, Seoul 08826, Republic of Korea}

\author[0000-0002-6810-1778]{Jaehyun Lee}
\email{jaehyun@kias.re.kr}
\affiliation{Korea Institute for Advanced Study, 85 Hoegi-ro, Dongdaemun-gu, Seoul 02455, Republic of Korea}
\affiliation{Korea Astronomy and Space Science Institute, 776 Daedeokdae-ro, Yuseong-gu, Daejeon 34055, Republic of Korea}
\affiliation{Department of Physics and Astronomy, Seoul National University, 1 Gwanak-ro, Gwanak-gu, Seoul 08826, Republic of Korea}

\author[0000-0002-1125-7384]{Aaron Meisner}
\email{aaron.meisner@noirlab.edu}
\affiliation{NSF NOIRLab, 950 N. Cherry Ave., Tucson, AZ 85719, USA}

\author{Ramon Miquel}
\email{rmiquel@ifae.es}
\affiliation{Instituci\'{o} Catalana de Recerca i Estudis Avan\c{c}ats, Passeig de Llu\'{\i}s Companys, 23, 08010 Barcelona, Spain}
\affiliation{Institut de F\'{i}sica d’Altes Energies (IFAE), The Barcelona Institute of Science and Technology, Edifici Cn, Campus UAB, 08193, Bellaterra (Barcelona), Spain}

\author[0000-0002-2733-4559]{John Moustakas}
\email{jmoustakas@siena.edu}
\affiliation{Department of Physics and Astronomy, Siena University, 515 Loudon Road, Loudonville, NY 12211, USA}

\author[0000-0001-9070-3102]{Seshadri Nadathur}
\email{seshadri.nadathur@port.ac.uk}
\affiliation{Institute of Cosmology and Gravitation, University of Portsmouth, Dennis Sciama Building, Portsmouth, PO1 3FX, UK}

\author[0000-0002-0905-342X]{Gautam Nagaraj}
\email{gautam.nagaraj@epfl.ch}
\affiliation{Laboratoire d'Astrophysique, EPFL, 1015 Lausanne, Switzerland}

\author[0000-0002-7356-0629]{Julie Nantais}
\email{julie.nantais@unab.cl}
\affiliation{Universidad Andres Bello, Facultad de Ciencias Exactas, Departamento de Fisica y Astronomia, Instituto de Astrofisica, Fernandez Concha 700, Las Condes, Santiago RM, Chile}

\author[0000-0001-9850-9419]{Nelson Padilla}
\email{n.d.padilla@gmail.com}
\affiliation{Instituto de Astronomía Teórica y Experimental (IATE), CONICET-UNC, Laprida 854, X500BGR, Córdoba, Argentina}

\author[0000-0001-9521-6397]{Changbom Park}
\email{cbp@kias.re.kr}
\affiliation{Korea Institute for Advanced Study, 85 Hoegi-ro, Dongdaemun-gu, Seoul 02455, Republic of Korea}

\author[0000-0002-0644-5727]{Will Percival}
\email{will.percival@uwaterloo.ca}
\affiliation{Department of Physics and Astronomy, University of Waterloo, 200 University Ave W, Waterloo, ON N2L 3G1, Canada}
\affiliation{Perimeter Institute for Theoretical Physics, 31 Caroline St. North, Waterloo, ON N2L 2Y5, Canada}
\affiliation{Waterloo Centre for Astrophysics, University of Waterloo, 200 University Ave W, Waterloo, ON N2L 3G1, Canada}

\author[0000-0001-7145-8674]{Francisco Prada}
\email{fprada@iaa.es}
\affiliation{Instituto de Astrof\'{i}sica de Andaluc\'{i}a (CSIC), Glorieta de la Astronom\'{i}a, s/n, E-18008 Granada, Spain}

\author[0000-0001-6979-0125]{Ignasi Pérez-Ràfols}
\email{ignasi.perez.rafols@upc.edu}
\affiliation{Departament de F\'isica, EEBE, Universitat Polit\`ecnica de Catalunya, c/Eduard Maristany 10, 08930 Barcelona, Spain}

\author{Graziano Rossi}
\email{graziano@sejong.ac.kr}
\affiliation{Department of Physics and Astronomy, Sejong University, 209 Neungdong-ro, Gwangjin-gu, Seoul 05006, Republic of Korea}

\author[0000-0002-9646-8198]{Eusebio Sanchez}
\email{eusebio.sanchez@ciemat.es}
\affiliation{CIEMAT, Avenida Complutense 40, E-28040 Madrid, Spain}

\author[0000-0002-3461-0320]{Joseph Harry Silber}
\email{jhsilber@lbl.gov}
\affiliation{Lawrence Berkeley National Laboratory, 1 Cyclotron Road, Berkeley, CA 94720, USA}

\author[0000-0002-4362-4070]{Hyunmi Song}
\email{hmsong@cnu.ac.kr}
\affiliation{Department of Astronomy and Space Science, Chungnam National University, 99 Daehak-ro, Yuseong-gu, Daejeon, 34134, Republic of Korea}

\author{David Sprayberry}
\email{david.sprayberry@noirlab.edu}
\affiliation{NSF NOIRLab, 950 N. Cherry Ave., Tucson, AZ 85719, USA}

\author[0000-0003-1704-0781]{Gregory Tarlé}
\email{gtarle@umich.edu}
\affiliation{University of Michigan, 500 S. State Street, Ann Arbor, MI 48109, USA}

\author[0000-0001-5567-1301]{Francisco Valdes}
\email{frank.valdes@noirlab.edu}
\affiliation{NSF NOIRLab, 950 N. Cherry Ave., Tucson, AZ 85719, USA}

\author[0000-0003-3078-2763]{Yujin Yang}
\email{yyang@kasi.re.kr}
\affiliation{Korea Astronomy and Space Science Institute, 776 Daedeokdae-ro, Yuseong-gu, Daejeon 34055, Republic of Korea}

\author[0000-0001-6047-8469]{Ann Zabludoff}
\email{aiz@arizona.edu}
\affiliation{Steward Observatory, University of Arizona, 933 North Cherry Avenue, Tucson, AZ 85721, USA}

\author[0000-0002-6684-3997]{Hu Zou}
\email{zouhu@nao.cas.cn}
\affiliation{National Astronomical Observatories, Chinese Academy of Sciences, A20 Datun Road, Chaoyang District, Beijing, 100101, P.~R.~China}
\begin{abstract}

We present a detailed study of the 3D morphology of two extended associations of multiple protoclusters at $z=3.1$. These protocluster `complexes', designated COSMOS-z3.1-A and COSMOS-z3.1-C, are the most prominent overdensities of $z=3.1$ Ly$\alpha$ emitters (LAEs) identified in the COSMOS field by the One-hundred-deg$^2$ DECam Imaging in Narrowbands (ODIN) survey. 
These protocluster complexes have been followed up with extensive spectroscopy from Keck, Gemini, and DESI. Using a probabilistic method that combines photometrically selected and spectroscopically confirmed LAEs, we reconstruct the 3D structure of these complexes on scales of $\approx$50 cMpc.
We validate our reconstruction method using the IllustrisTNG300-1 cosmological hydrodynamical simulation and show that it consistently outperforms approaches relying solely on spectroscopic data. The resulting 3D maps reveal that both complexes are irregular and elongated along a single axis, emphasizing the impact of sightline on our perception of structure morphology. The complexes consist of multiple density peaks, ten in COSMOS-z3.1-A and four in COSMOS-z3.1-C. The former is confirmed to be a proto-supercluster, similar to {\it Hyperion} at $z=2.4$ but observed at an even earlier epoch. Multiple `tails' connected to the cores of the density peaks are seen, likely representing cosmic filaments feeding into these extremely overdense regions. The 3D reconstructions further provide strong evidence that Ly$\alpha$ blobs preferentially reside in the outskirts of the highest density regions. Descendant mass estimates of the density peaks suggest that COSMOS-z3.1-A and COSMOS-z3.1-C will evolve to become ultra-massive structures by $z=0$, with total masses $\log(M/M_\odot) \gtrsim 15.3$, exceeding that of Coma.

\end{abstract}

\section{Introduction}\label{sec:intro}





In the hierarchical theory of structure formation, massive structures form through gravitationally driven growth and collapse of density fluctuations in the primordial Universe. In this paradigm, the most massive structures - namely, clusters of galaxies with mass $\gtrsim 10^{14} M_\odot$ - grow and evolve by continuously accreting matter from the surrounding cosmic web. At $z \gtrsim 2$, the progenitors of these clusters, or `protoclusters', accrete cold gas along the connected filaments \citep[cold accretion:][]{dekel09} to support vigorous star-formation and AGN activity \citep[e.g.,][]{Casey2015, Umehata2015,wang16,Oteo2018,Harikane2019,Lemaux2022}. Understanding the formation and evolution of protoclusters, therefore, requires studying them within the broader context of the cosmic web.


In the past decade, wide-field imaging surveys have been successful in identifying (comparatively) large samples of protoclusters \citep[e.g.,][]{Chiang2014, Toshikawa2018, Harikane2019, Ramakrishnan2024, Takeda2024, Hung2025} at $z \sim 2 - 7$. These surveys have granted valuable insight into the formation of (proto) cluster galaxies, demonstrating the rapid and early build-up of massive galaxies within these environments \citep[e.g.][]{Ito2020,Toshikawa2024}. 

Recently, in \citet{Ramakrishnan2024,Ramakrishnan2025}, we reported the detection of $\sim$ 150 protoclusters at $z~=~2.4$ and 3.1 using data from the One-hundred-deg$^2$ DECam Imaging in Narrowbands \citep[ODIN,][]{Lee2024} survey. ODIN uses three narrowband filters to identify Ly$\alpha$-emitting galaxies (LAEs) at $z~=~2.4$, 3.1, and 4.5 \citep{Firestone2024}, thereby tracing the large-scale structure in thin, well-defined cosmic slices ($\Delta z = 0.06 - 0.08$) over contiguous areas of $\sim$10-20~deg$^2$ in each field. The structures revealed by ODIN are diverse in size and strikingly non-symmetric in morphology, in good alignment with expectations from cosmological hydrodynamical simulations \citep{Ramakrishnan2024}. They frequently lie at the intersection of multiple cosmic web filaments also traced by LAEs, in accordance with the hierarchical theory of structure formation. 

ODIN protoclusters are also shown to be closely associated with rare, extended and luminous Ly$\alpha$ `blobs' \citep[LABs;][B.~Moon et al.\ in prep.]{Ramakrishnan2023}. These LABs present a rare opportunity to witness the gas of the intergalactic medium in emission. Their bright luminosity requires an abundant source of ionizing photons, with proposed powering mechanisms including vigorous star formation \citep{cen13,Geach2016, Ao2017}, AGN or QSO activity \citep{Dey2005, Geach2009, Yang2014a, Cai2017}, and gravitational cooling \citep{Fardal2001, Rosdahl2012, Daddi2021, Fabrizio22}. The observed close connection of LABs with ODIN-selected protoclusters hints at the various mechanisms shaping galaxy evolution in these massive structures.

Yet, the conclusions that can be drawn regarding the association of the ODIN protoclusters with cosmic filaments, LABs, and rare galaxy populations are fundamentally limited by the lack of line-of-sight information. Though the redshift uncertainties associated with ODIN LAEs and LABs are far smaller than those associated with, e.g., Lyman break galaxies ($\Delta z \sim 0.3-0.4$), they are still far larger than the typical size of a protocluster \citep[$\Delta z \sim 0.01 - 0.02$,][]{Chiang2013}. To place LABs and rare galaxies within the ODIN structures, as well as to explore their connection with the surrounding cosmic web, a 3D perspective is necessary.

To this end, we present in this work extensive spectroscopy obtained for the two most prominent ODIN protoclusters identified in the 9~deg$^2$ extended COSMOS field. In \citet{Ramakrishnan2023}, we referred to them as `Complex~A' and `Complex~C' and provided extensive discussion of their morphological characteristics based on the 2D overdensity map. In this paper, we revise these designations as `COSMOS-z3.1-A' and `COSMOS-z3.1-C', respectively, to specify the field and redshift range as there is a need to distinguish them from other structures of this magnitude identified in other ODIN fields. 
{These spectroscopic data show the large-scale structure in and around these protoclusters at spatial scales out to 50--100~cMpc, revealing multiple closely-spaced overdensities which are likely connected via cosmic filaments.}

This paper is organized as follows. In Section~\ref{sec:data}, we describe the imaging and spectroscopy data used for the analyses. Section~\ref{sec:reconstruction} outlines the methodology we employed to reconstruct the three-dimensional matter distribution from spectroscopy. We validate our method by creating mock data from cosmic structures identified in cosmological hydrodynamic simulations. The 3D structure maps of the two protoclusters are presented in Section~\ref{sec:odin_structures}. In Section \ref{sec:discussion}, we discuss the conclusions which can be drawn from our 3D maps, most notably demonstrating that COSMOS-z3.1-A is a structure similar to {\it Hyperion}, the only known proto-supercluster at $z=2.45$ \citep{Cucciati2018}. Finally, in Section~\ref{sec:outlook}, we discuss the optimal strategy for future spectroscopy to further explore the large-scale structure of more protoclusters. We adopt a $\Lambda$CDM concordance cosmology with $h = 0.7$, $\Omega_m = 0.27$, and $\Omega_\Lambda = 0.73$ and use comoving distance scales unless noted otherwise. 

\section{Data} \label{data}\label{sec:data}

\subsection{Keck/DEIMOS Spectroscopy} \label{subsec:keck}

\begin{deluxetable*}{ccccccc}
\begin{enumerate}
    \item[] \tablecaption{Summary of spectroscopic observations \label{tab:spec_obs}}
\end{enumerate}
\tablehead{\colhead{Mask Name} & \colhead{Center R.A.} & \colhead{Center Decl.} & \colhead{PA} & \colhead{$t_{\rm exp}$ (s)} & \colhead{$N_{\rm LAE}^\dagger$} & \colhead{$N_{\rm LAB}^\ddagger$} 
}
\startdata
\hline
\multicolumn{7}{c}{\bf{Keck/DEIMOS}} \\
\hline
COSMOS-z3.1-A\_1 & $10^h01^m56^s.76$ & $+2^d28^m39^s.0$ & 113$^\circ$ & 10200 & 24 (27) & 0 (0)\\
COSMOS-z3.1-A\_2 &$10^h02^m53^s.02$ & $+2^d53^m46^s.9$ & 30$^\circ$ & 5280 & 16 (21) & 1 (1)\\
COSMOS-z3.1-A\_3 &$10^h03^m04^s.61$ & $+2^d33^m17^s.2$ & 115$^\circ$ & 4800 & 19 (26) & 0 (0) \\
COSMOS-z3.1-A\_4 &$10^h02^m33^s.56$ & $+2^d39^m30^s.1$ & 10$^\circ$ & 5400 & 17 (24) & 3 (4) \\
COSMOS-z3.1-A\_5 &$10^h04^m13^s.53$ & $+2^d39^m17^s.8$ & 60$^\circ$ & 7500 & 12 (20) & 2 (2) \\
COSMOS-z3.1-A\_6 &$10^h03^m52^s.05$ & $+2^d29^m41^s.0$ & 120$^\circ$ & 7200 & 8 (20) & 0 (0) \\
COSMOS-z3.1-A\_7 &$10^h03^m23^s.42$ & $+2^d42^m18^s.53$ & 0.4$^\circ$ & 8400 & 15 (23) & 1 (1) \\
\hline
\multicolumn{7}{c}{{\bf Gemini/GMOS}} \\
\hline
COSMOS-z3.1-C\_1 & $9^h54^m53^s.72$ & $+2^d46^m47^s.4$ & 70$^\circ$ & 7000 & 6 (8) & 1 (1)\\
COSMOS-z3.1-C\_2 & $9^h55^m19^s.49$ & $+2^d48^m48^s.8$ & 160$^\circ$ & 7000 & 6 (11) & 0 (0)\\
COSMOS-z3.1-C\_3 &$9^h56^m01^s.32$ & $+2^d20^m52^s.7$ & 100$^\circ$ & 7200 & 6 (8) & 1 (2) \\
COSMOS-z3.1-C\_4 & $9^h55^m13^s.27$ & $+2^d45^m31^s.1$ & 180$^\circ$ & 7720 & 9 (11) & 0 (0) \\
\hline
\multicolumn{7}{c}{{\bf DESI}} \\
\hline
COSMOS-z3.1 & $10^h00^m24^s.00$ & $+2.0^d10^m55^2$ & -- & -- & 1285 (1422) & -- \\
\enddata
\tablecomments{The pointing center of the eleven multi-object masks is listed together with the position angle (PA) and exposure time. The number of confirmed (targeted) LAE and LAB candidates on each mask is indicated in the column marked by $\dagger$ and $\ddagger$, respectively.}
\end{deluxetable*}

COSMOS-z3.1-A was followed up with the Keck II/DEIMOS instrument \citep{DEIMOS_ref} through three dedicated programs (program IDs U151, U222, and R370). The primary targets were ODIN $N501$ LAEs \citep{Firestone2024} and LABs \citep{Ramakrishnan2023,Moon2026}, with $u$-dropout galaxies used as filler targets. The latter were chosen using $u$-band imaging from CLAUDS \citep{Sawicki2019} and $gri$ data from HSC-SSP \citep{Aihara2018a,Aihara2019}. We detected sources in the $i$ band, which has the best seeing \citep{Aihara2018a,Aihara2018b}, and selected $u$-band dropouts using the following criteria:
\begin{equation}
\begin{split}
    u - g > 1.2 \\
    0.1 < g - r < 0.7 \\
    u - g > 2.5(g - r) + 0.5 \\
    r - i < 0.3
\end{split}    
\end{equation}
These criteria were calibrated to select galaxies with a median redshift of $z = 3.1$ using photometric redshifts from the COSMOS2020 \citep{Weaver2022:COSMOS2020} catalog. COSMOS2020 makes use of multiband data spanning the UV to mid-IR to determine the photometric redshifts of galaxies, achieving an accuracy of $\Delta z \lesssim 0.025(1+z)$.  

A total of seven masks were observed with exposure times ranging from 5400 - 10200 s per mask, covering 161 LAEs and 8 LABs, as summarized in Table \ref{tab:spec_obs}. We utilized the 900ZD grating centered at a wavelength of 5500~\AA\ with the GG400 blocking filter, resulting in a wavelength coverage of $\sim$4000~\AA\--7200~\AA. The slit width was 1\arcsec, corresponding to a 
resolving power of $R\sim1964$. We acquired arcs for wavelength calibration using the Ne, Ar, Kr, and Xe lamps as well as the Cd, Zn, and Hg lamps, with the latter being required as our wavelength of interest (i.e., that of the Ly$\alpha$ line) was towards the blue end of the observable wavelengths. Data reduction was carried out with the \textsc{PypeIt} software \citep{Prochaska2020}, using the default settings for the Keck/DEIMOS spectrograph. We visually inspected the extracted 2D spectra and coadded 1D spectra to search for evidence of the Ly$\alpha$ line. In all cases where a Ly$\alpha$ line was detected, we set the redshift of the source to correspond to the peak of the line. In total, we confirmed 111 out of 161 LAEs and 7 out of 8 LABs. 

Given the exposure time, we did not observe any continuum emission for the majority of the $u$-dropout galaxies, and they remain unconfirmed. However, 9 $u$-band dropout galaxies displayed a single strong emission line in the wavelength range 4700~\AA--5200~\AA, which we assumed to be Ly$\alpha$ given their wavelength and the absence of other detectable emission lines.

\newcommand{\civlamb}{\ion{C}{4}\,$\lambda$1548,\,1550\xspace}
\newcommand{\ciilamb}{\ion{C}{2}\,$\lambda$1335\xspace}
\newcommand{\heiilamb}{\ion{He}{2}\,$\lambda$1640\xspace}

\begin{figure*}
    \centering
    \includegraphics[width=0.9\linewidth]{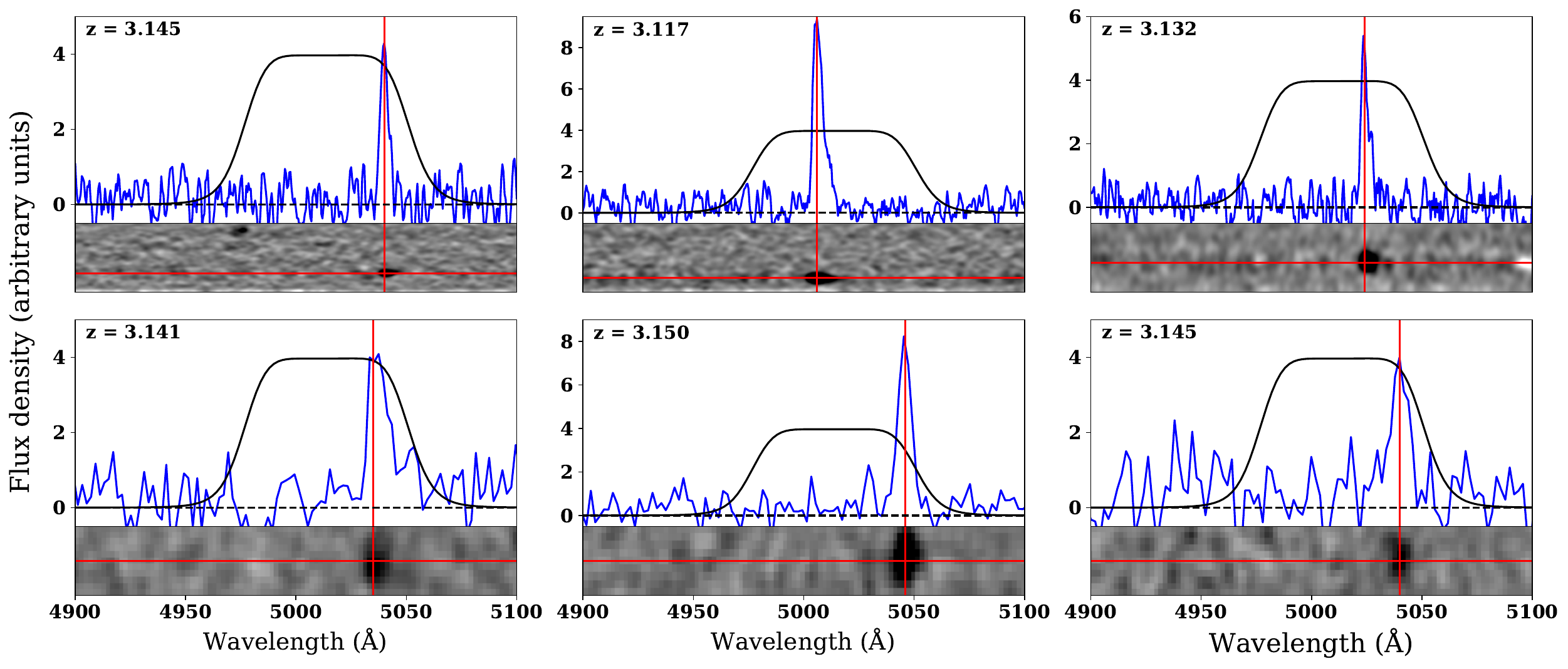}
    \caption{Examples of the spectra of LAEs confirmed with Keck II/DEIMOS (top row) and Gemini/GMOS (bottom row). The black curve shows the $N501$ filter transmission function, while the peak of the Ly$\alpha$ line, used to infer the redshift, is indicated by a vertical red line. The 2D spectra (shown in the bottom of each panel) are slightly smoothed to better distinguish the presence of line emission. Red horizontal lines in the 2D spectra indicate the position of the source within the slit, slightly offset for clarity.}
    \label{fig:spectra}
\end{figure*}

\begin{figure*}
    \centering
    \includegraphics[width=0.8\linewidth]{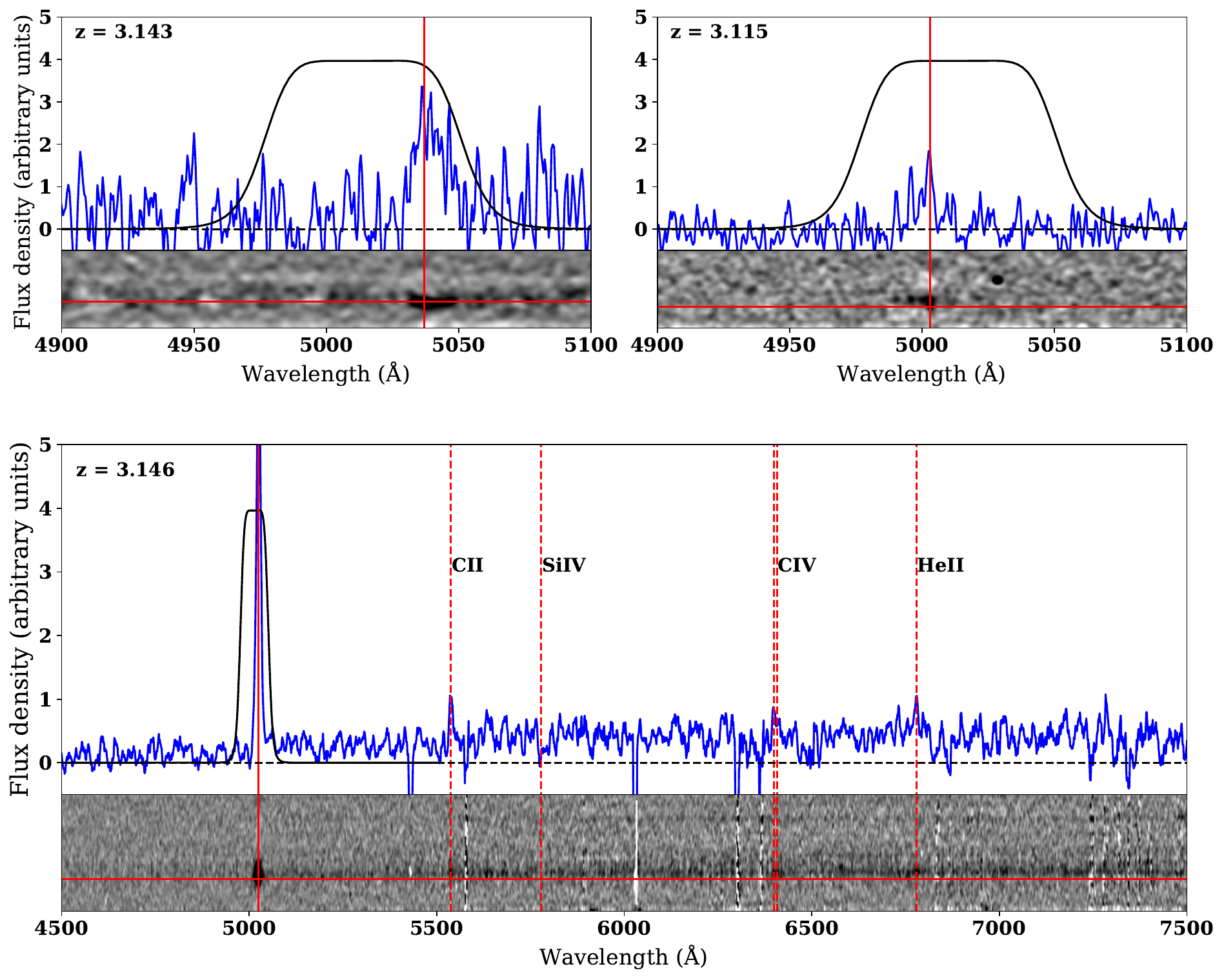}    
    \caption{As Figure~\ref{fig:spectra}, but for LABs confirmed with DEIMOS. The Ly$\alpha$ line of LABs is, in general, wider than that of LAEs. The more luminous LABs show clear evidence of AGN origin, with a bright continuum and multiple observed emission lines (as shown in the bottom panel).}
    \label{fig:lab_spec}
\end{figure*}

In the top row of Figure \ref{fig:spectra}, we show example spectra of the LAEs confirmed with DEIMOS. The Ly$\alpha$ line is easily identifiable through its characteristic asymmetry, with the red wing being more pronounced than the blue. In Figure \ref{fig:lab_spec}, we show the spectra of some of the confirmed LABs. In general, the Ly$\alpha$ line in the LAB spectra is broader than that of the LAEs. One of the confirmed LABs is clearly an AGN, displaying broad Ly$\alpha$ emission as well as visible \ciilamb, \civlamb and \heiilamb lines (the bottom panel of Figure \ref{fig:lab_spec}).


\subsection{Gemini/GMOS Spectroscopy} \label{subsec:gemini}

COSMOS-z3.1-C was followed up with the Gemini/GMOS instrument \citep{GMOS_ref} in the 2023A and 2025A semesters with four masks targeting ODIN $N501$ LAEs and LABs.
We used the B600 grating to obtain moderate sensitivity over a wide spectral range, facilitating the detection of Ly$\alpha$ as well as other UV emission lines (e.g., \civlamb and \heiilamb).
The four masks covered a total of 38 LAEs and 3 LABs (see Table \ref{tab:spec_obs}).
For LAEs, we adopted a 1\arcsec\ slit width corresponding to $R\sim844$ and 7\arcsec\ to 12\arcsec\ lengths to get enough empty sky region for a sky subtraction,
while for LABs we utilized slits with 1\arcsec\ and 2\arcsec\ widths ($R=844$ and 422, respectively) and 20\arcsec\ to 30\arcsec\ lengths.
We adopted a wavelength dither of 100--200~\AA\ to avoid the observed Ly$\alpha$ emission line falling into the chip gaps.

Observations were carried out with on source time $\sim$7000~s for each mask with $\sim$1000~s single exposures. We reduced the data with the Gemini IRAF package and confirmed 27 out of 38 LAEs through their narrow Ly$\alpha$ emissions. Two out of three LABs are confirmed in the same way as LAEs. The others showed no emission lines due to faint narrowband magnitudes. 

In the bottom row of Figure~\ref{fig:spectra}, we show example spectra of LAEs confirmed with Gemini/GMOS. {Further results including the spectra of LABs confirmed with Gemini/GMOS are presented in \citet{Moon2026}.}

\begin{figure*}
    \centering
    \includegraphics[width=0.85\linewidth]{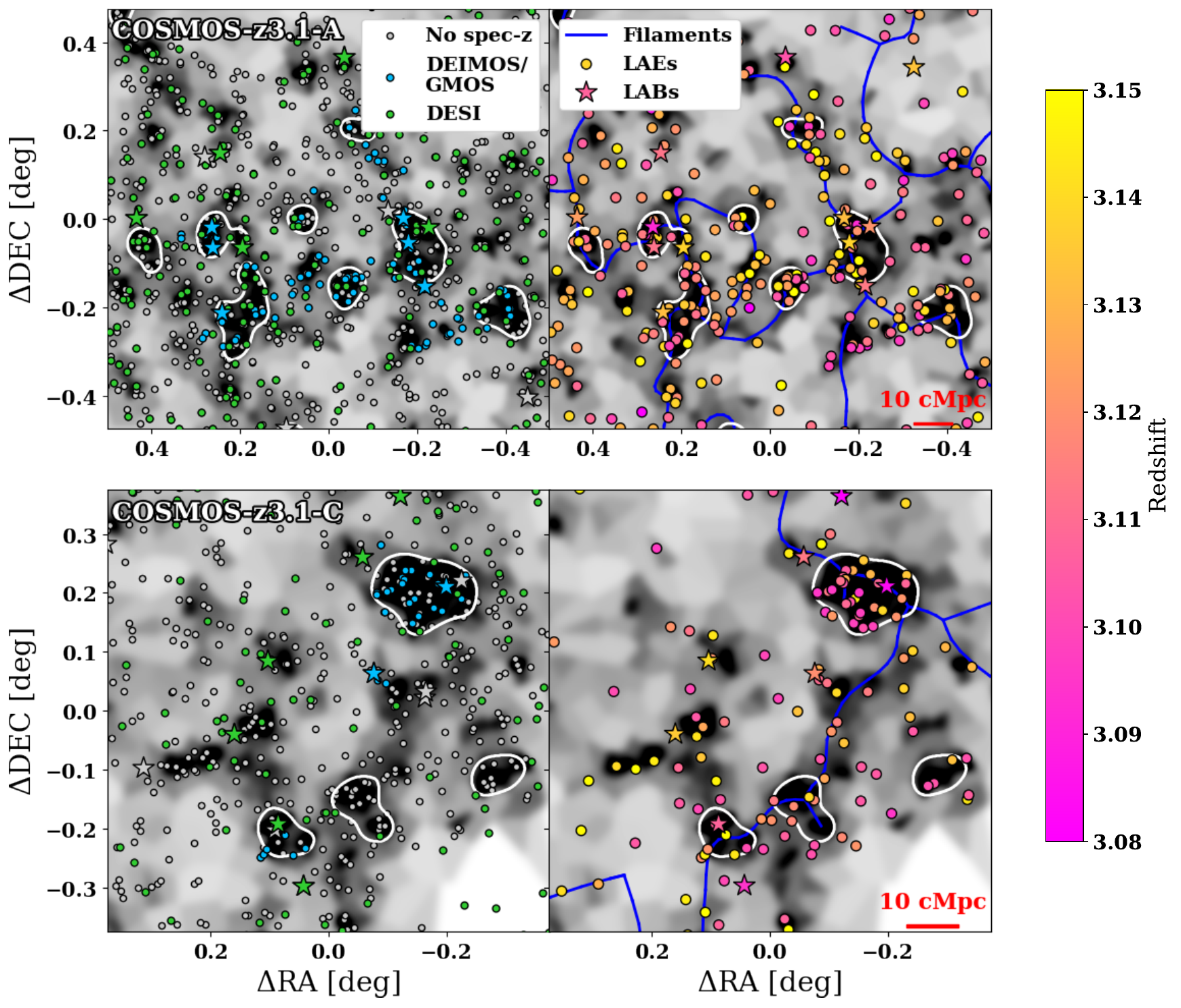}
    \caption{Spectroscopically confirmed sources for COSMOS-z3.1-A (top) and COSMOS-z3.1-C (bottom), overlaid on the 2D LAE surface density map. White contours show the boundaries of the protocluster candidates identified from the 2D surface density map (i.e. as overdensities of photometrically selected LAEs). 
    In the left-hand panels, all identified LAEs (dots) and LABs (stars) are shown, with blue symbols indicating objects confirmed via targeted DEIMOS/GMOS spectroscopy, green indicating those confirmed via wide-field DESI spectroscopy, and grey those which remain unconfirmed. In the right-hand panels, all confirmed LAEs and LABs are shown color-coded by redshift.{Blue lines show filaments identified from the distribution of photometrically selected LAEs with the Discrete Persistent Structure Extractor \citep[DisPerSE,][]{Sousbie2011a,Sousbie2011b} software.}} 
    \label{fig:spec_sources}
\end{figure*}

\subsection{DESI Spectroscopy} \label{subsec:desi}



The targeted spectroscopic follow-up is supplemented by wide area spectroscopy of ODIN LAEs obtained with the Dark Energy Spectroscopic Instrument \citep[DESI;][]{DESI_ref}. DESI is a wide-field, fiber-fed spectrograph mounted on the Mayall 4m telescope at Kitt Peak National Observatory \citep{DESI2022:instrumentation}, currently carrying out a long-term survey of $\sim$ 17,000 deg$^2$ of the sky \citep{DESI2023:survey,DESI2025:DR1} to place constraints on the nature of dark energy \citep{DESI2025:cosmology,DESI2025:BAO}. 

DESI is capable of obtaining spectra of up to $\sim$5000 sources per pointing \citep{DESI2016:instrument_design,DESI2024:optical_corrector,DESI2024:fiber_system}. The DESI-ODIN observations were hence aimed at confirming several thousand LAEs at $z=2.4$ and 3.1 over the full area of the COSMOS and XMM-LSS fields. The redshifts of these LAEs were determined via a detailed visual inspection program \citep{Pinarski2026} of the spectra reduced via the DESI pipeline \citep{DESI2023:specz_pipeline}. Details of the DESI target selection, observations, and data reduction will be presented in a future paper (A.~Dey et al.\ in prep.).

In Figure \ref{fig:spec_sources}, we show the spectroscopically confirmed sources for our two complexes. The greyscale background in all panels indicates the LAE {surface density} map computed by the 2D Voronoi tessellation \citep[see][for more detail]{Ramakrishnan2024}, while white contours show the 2D overdensities selected as protoclusters. The left panels indicate the sky positions of all photometric LAE and LAB candidates. Sources confirmed via targeted Keck/DEIMOS or Gemini/GMOS follow-up are shown in blue, those confirmed via DESI observations in green, and unconfirmed sources in grey. In the right panels, we show all spectroscopically confirmed LAEs and LABs, color-coded by redshift. The fraction of spectroscopically confirmed LAEs within the two complexes is $\sim$~0.2--0.4 on average. 

\section{Reconstructing the large-scale structure in 3D} \label{sec:reconstruction}

To place the formation and evolution of galaxies in the context of large-scale structure (LSS), a robust determination of the three-dimensional positions for individual galaxies is necessary. The typical diameter of a protocluster is $\sim$10--15~cMpc \citep{Chiang2013}; distinguishing galaxies residing near the core of an overdensity from those at the outskirts (5--7~cMpc from the center) or well outside it ($\gtrsim$7--10~cMpc) thus requires spatial resolution better than $\approx$2--3~cMpc. One unit comoving megaparsec (1~cMpc) corresponds to $\Delta z=0.001$ at $z=3.12$. This redshift resolution can only be achieved through spectroscopy.


When examining the left and right panels of Figure \ref{fig:spec_sources}, we make two observations. First, many galaxies within a specific overdensity share similar redshifts ($\Delta z\lesssim 0.03$, or 30~cMpc at $z=3.1$), indicating their physical proximity. It is also apparent that the surface density of photometric candidates (grey sources) significantly surpasses that of spectroscopic sources. Although the redshifts of photometric sources are not known individually, they are expected to be confined to a narrow range. The wavelength range at which the $N501$ transmission drops to 50\% of its maximum corresponds to $z=3.093-3.155$, which aligns with the redshift range occupied by spectroscopically confirmed LAEs. 

In this work, we explore ways to probabilistically reconstruct the LSS by combining the spectroscopic information and the overall redshift distribution. This is similar to and inspired by the method employed by \citet{Cucciati2018,Lemaux2018,Hung2020}. 
The key difference in our case is that nearly all galaxies lie in a very narrow redshift range, which may be an advantage. After we describe the adopted procedure in Section~\ref{subsec:method}, we evaluate the efficacy of the method using the IllustrisTNG300-1 \citep[TNG300 hereafter;][]{Pillepich2018a,Pillepich2018b,Nelson2019} simulation in Section~\ref{subsec:validation}. TNG300 offers the largest volume (302.6~cMpc on a side) of all available simulations of the IllustrisTNG suite, with mass resolution of $1.1\times 10^7 M_\odot$ and $5.9\times 10^7 M_\odot$ for baryons and dark matter, respectively. Stellar mass values are assigned to each collapsed dark matter halo.

Before discussing our methodology, we note two caveats. First, the peculiar motion of a galaxy will affect its observed spectroscopic redshift, offsetting it from the true cosmological redshift. This gives rise to redshift-space distortions in the reconstructed structures, causing them to be elongated \citep[`Finger-of-God' effect,][]{Jackson1972} or compressed \citep[`Pancake-of-God' or Kaiser effect,][]{kaiser87} along the redshift axis. Secondly, as a result of the resonant nature of the Ly$\alpha$ transition, the Ly$\alpha$ line is typically redshifted relative to the systemic velocity, with the magnitude of the velocity offset depending on factors such as the equivalent width and star-formation rate \citep[e.g.,][]{Erb2014, Muzahid2020}. In Appendix~\ref{appendix:peculiar_motion}, we show that these redshift uncertainties do not have a significant effect on our 3D reconstructions.

\subsection{Probabilistic reconstruction of the LSS}\label{subsec:method}

\begin{figure*}
    \centering
    \includegraphics[width=0.48\linewidth]{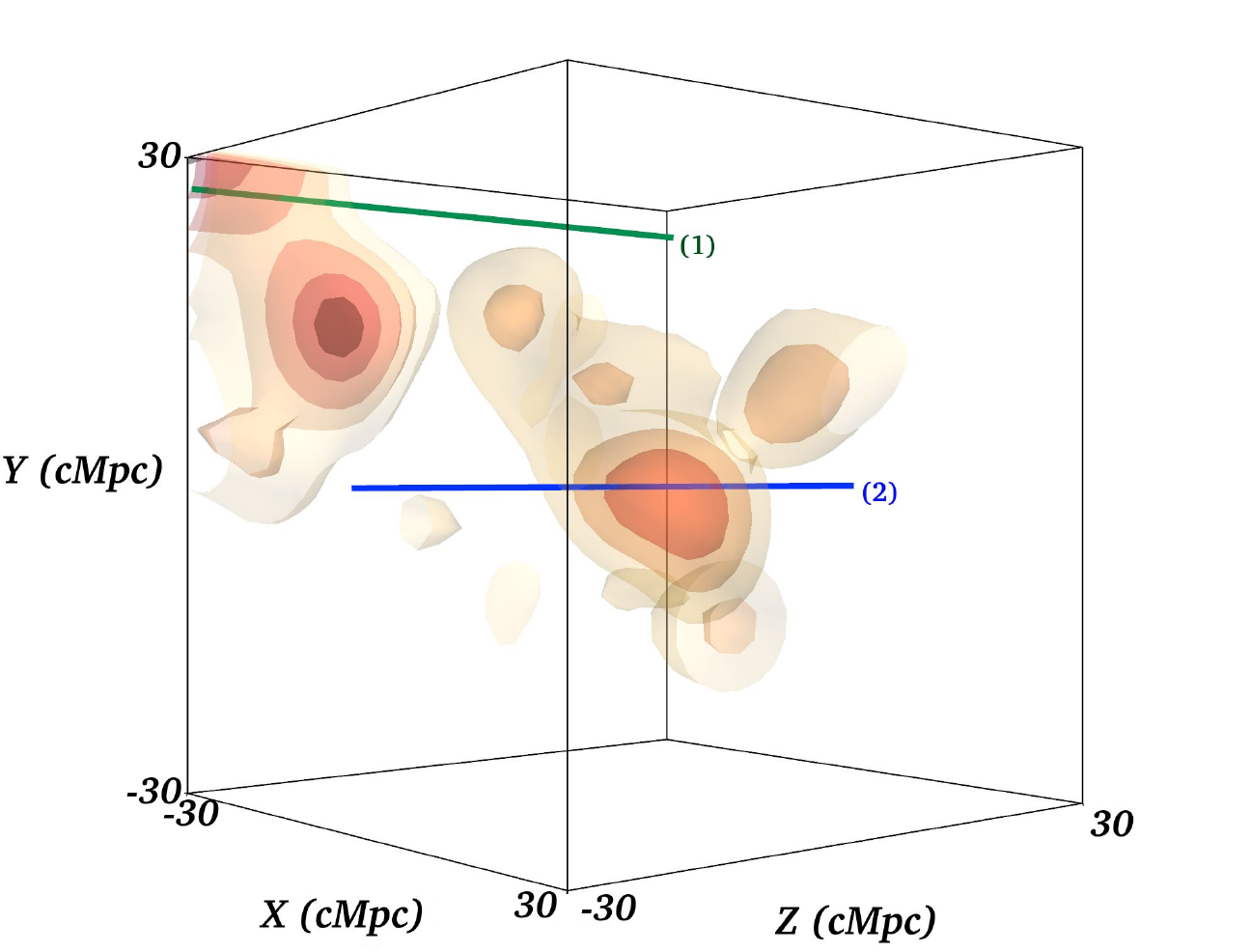}
    \includegraphics[width=0.48\linewidth]{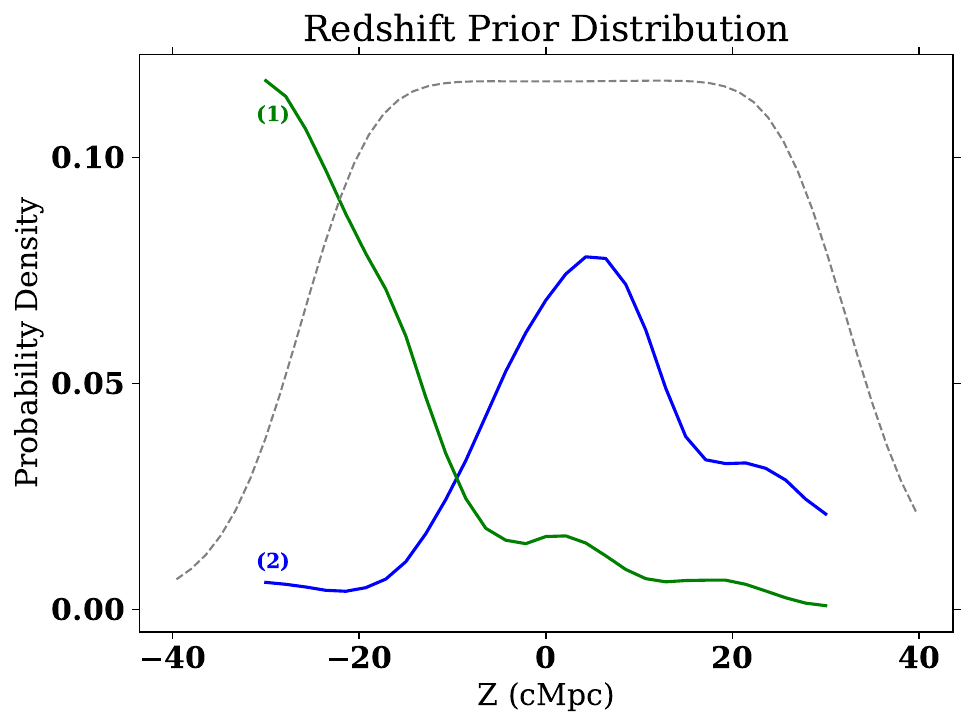}
    
    \caption{\textit{Left:} A 3D visualization of the smoothed priors showing the spatial distribution of galaxies, with two lines (green and blue) representing the line-of-sight at two randomly chosen $x$ and $y$ positions. Note that the third and fourth vertical lines from the left indicate the back face of the cubic volume under consideration. \textit{Right:} Probability distributions along the line-of-sight for each corresponding position (green and blue lines), showing the prior used to assign redshifts to the LAEs with projected positions close to those $x$ and $y$ coordinates. The $N501$ transmission curve is shown as a grey dashed line.}
    \label{fig:3d_prior}
\end{figure*}

Our methodology is built on the premise that, in a given sightline ${\mathcal S_i}$, the overall redshift distribution, $N(z,{\mathcal S_i})$, is similar to that of the spectroscopically confirmed LAEs, $N_{\rm spec}(z,{\mathcal S_i})$. Equating $N(z,{\mathcal S_i})$ and $N_{\rm spec}(z,{\mathcal S_i})$ in this manner implicitly assumes that: 1) the linear bias of spectroscopically confirmed and photometrically selected LAEs is similar; and 2) the contamination of the photometrically selected LAEs is minimal. These assumptions have been shown to be valid in the case of the ODIN data. The LAEs confirmed through our targeted spectroscopic follow-up are chosen at random from the photometrically selected LAEs falling within a mask. Furthermore, the bias values found for the full ODIN LAE samples at $z=2.4$ and 3.1 by \citet{herrera25} are consistent within the uncertainty with those found by \citet{White2024} for ODIN LAEs followed up with DESI. Likewise, our spectroscopic follow-up has shown that the ODIN LAE sample has a purity of $\gtrsim$ 95\% (E.~Pinarski et al.\ in prep.). We thus feel confident in assuming that $N(z,{\mathcal S_i}) \simeq N_{spec}(z,{\mathcal S_i})$.

Once $N(z,{\mathcal S_i})$ is known, we can use it as a prior and assign redshifts to {\it all} LAE candidates. Figure~\ref{fig:3d_prior} illustrates two different sightlines within a cosmic structure at $z=3$ in TNG300. Generally, the efficacy of this method is expected to improve with increasing fraction of spectroscopically confirmed sources. Additionally, the angular size of a unique sightline, ${\mathcal S_i}$, will also matter. If the size of a cell representing a sightline ${\mathcal S_i}$ is too small, the number of (spectroscopically) confirmed sources would be too small and fail to provide a robust statistic; similarly, if the cell size is too large compared to the angular extent of characteristic features in the LSS, the resultant redshift function will not accurately capture the true 3D density fluctuation.


We begin by creating a uniform 3D grid with a cell size of $2\times2\times 2$~cMpc$^3$ centered around each protocluster. This corresponds to 60--70\arcsec\ in angular directions and $\Delta z \approx 0.002$ in the line-of-sight direction. All spec-z sources are sorted into the grid, and the resultant distribution is smoothed using a 3D Gaussian kernel with a standard deviation of 2--3~cMpc. Both bin size and smoothing scale are chosen such that we can maintain good spatial resolution while avoiding artifacts created by sparse sampling and shot noise after extensive tests conducted on cosmic structures identified in simulations. In Appendix~\ref{appendix:priors}, we present the effect of different choices of binning and smoothing scales on the 3D model of COSMOS-z3.1-C. As a last step, we add to the smoothed 3D histogram a constant (which we choose to be 1\% of the maximum probability) uniformly to each bin, to prevent regions of zero probability. The final redshift prior for a sightline ${\mathcal S_i}$, $N(z,{\mathcal S_i})$, is taken to be the smoothed histogram in the $i$th cell. In the right panel of Figure~\ref{fig:3d_prior}, we show the redshift prior for two example sightlines converted into a probability density function in the line-of-sight direction.  


Utilizing sightline-specific redshift priors, we assign redshifts to all LAE candidates in and around the ODIN structures. In other words, an unconfirmed LAE candidate falling within the $k$th cell is assigned a redshift drawn according to the probability distribution $N_{\rm spec}(z,{\mathcal S_k})$. There are LAEs belonging in cells with no determination of $N_{\rm spec}(z,{\mathcal S_i})$. Such cells form $\sim$2\% of the cases in our data; this typically occurs in low-density regions where spectroscopic coverage is scant. In the absence of any other information, we use a redshift prior corresponding to the ODIN filter transmission (also shown in Figure~\ref{fig:3d_prior}) in these cases. Taking the wavelengths at which filter transmission is 50\% of its maximum \citep{Lee2024}, the corresponding front-to-back thickness is 59.9~cMpc at $z=3.12$. Having assigned redshifts to all LAE candidates, we obtain a 3D model of a protocluster in spatial resolution of 2~cMpc. The assignment process is repeated 500 times. By averaging over multiple realizations, we reduce bias and ensure a statistically robust reconstruction.

\begin{figure*}
    \centering
    \includegraphics[width=\linewidth]{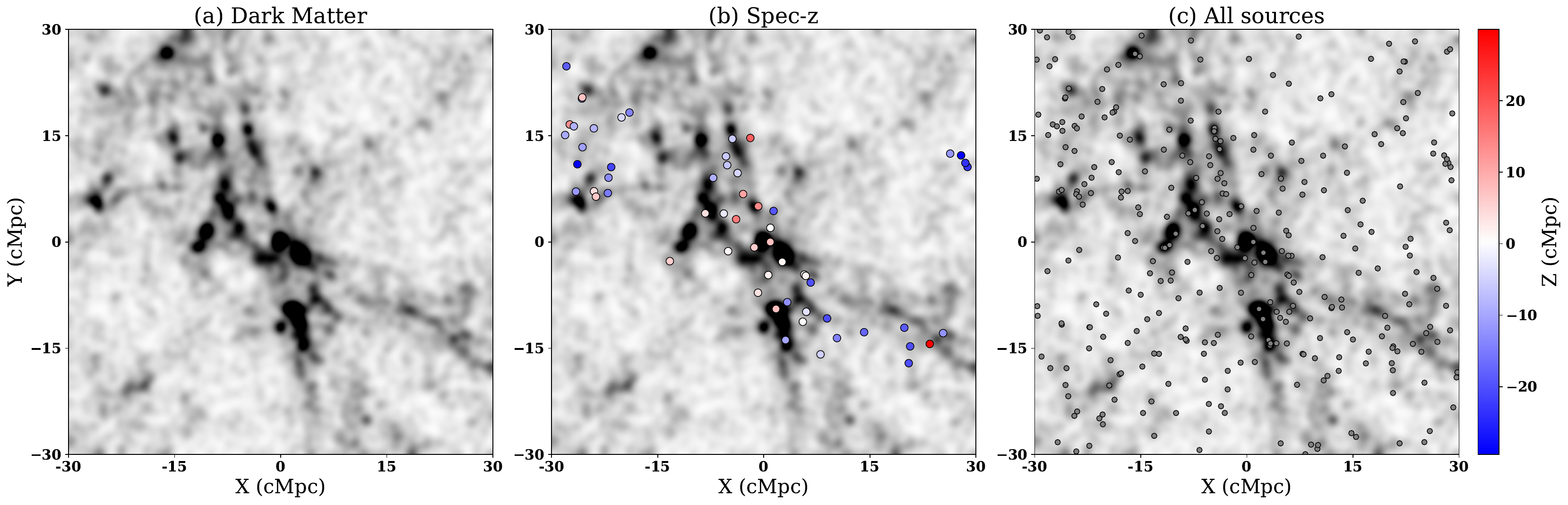}
    \includegraphics[width=0.95\linewidth]{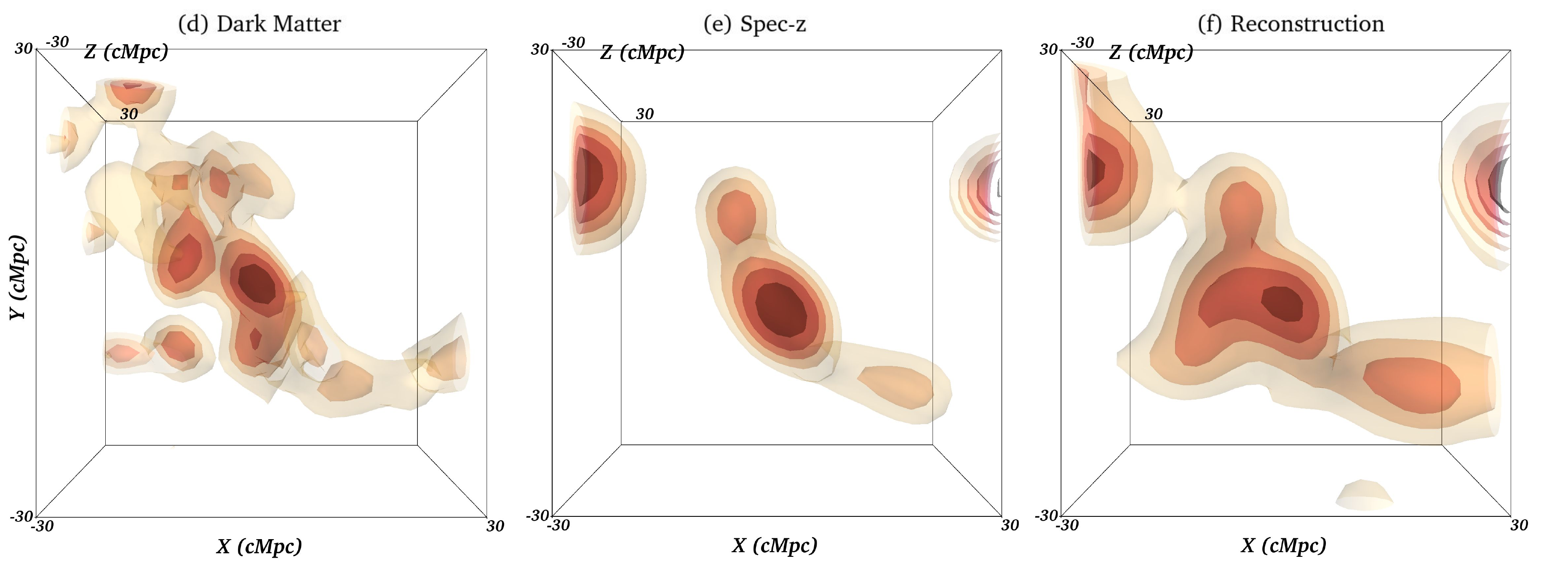}
    \caption{Comparison of the spatial distributions of LAEs and the underlying dark matter density in a 60x60x60 cMpc$^3$ region centered on the largest structure in TNG300 at $z~=~3$. \textit{Top:} (a) Projected dark matter density (smoothed for visualization) over a 60x60x20 cMpc$^3$ slab. (b) Same dark matter projection, overlaid with spec-z mock LAE sample, color-coded by redshift. (c) Same dark matter projection, overlaid with all mock LAE sources in the sample (without redshift). \textit{Bottom:}  (d) The underlying dark matter distribution in 3D, serving as a theoretical reference. (e) The spec-z mock LAE sample in 3D, used as priors. (f) The full probabilistic reconstruction, which more closely follows the dark matter distribution, particularly in the central overdense regions. This demonstrates that probabilistic redshift assignment improves the recovery of LSS compared to relying solely on spec-z sources.}
    \label{fig:tng300}
\end{figure*}

\subsection{Validation of 3D reconstruction methodology} \label{subsec:validation}


To evaluate the robustness of our 3D reconstruction methodology, we utilize the TNG300 simulation. In creating mock LAE samples from TNG300, we require that they match key characteristics of the ODIN LAE sample. The full description of the procedure is given in \citet[][Section~2.2]{Ramakrishnan2024}, and we only briefly summarize it here. 

First, we define (60~cMpc)$^3$ volumes of the $z=3$ TNG300 snapshot, centered on the barycenters of several massive cosmic structures that will evolve to Coma-like galaxy clusters by $z=0$ \citep{andrews2024}. This is motivated by the present-day mass estimates of the two ODIN structures, which place them high on the list of cosmic structures ranked by mass. Our descendant mass estimate and calibration will be discussed in Section~\ref{sec:odin_structures}. The 60 cMpc size in the line-of-sight direction roughly corresponds to the full-width-at-half-maximum of the $N501$ filter. Second, to ensure that the clustering properties of mock LAEs are similar to those of real LAEs, we require that their stellar masses obey a lognormal distribution: i.e., $\log (M_\ast/M_\odot)$ follows a normal distribution with mean and standard deviation $(\mu, \sigma)=(8.75,0.75)$ \citep[see, e.g.,][]{Hagen2014,Vargas2014}. Third, we randomly select a subset of halos selected by the first and the second criteria, to match $\bar{\Sigma}_{\rm LAE}$, the mean surface density of real ODIN LAEs. Both $\bar{\Sigma}_{\rm LAE}$ and the selected fraction are determined away from any large overdensities.

The final step is to assign a subset of this simulated LAE sample to be mock spec-z sources. To mimic the nature of our spectroscopic observations (specifically targeting regions of high LAE overdensities: see Figure~\ref{fig:spec_sources}), spec-z designations are given only in regions with the local surface density of mock photometric LAEs within the top 25\% in the (60~cMpc)$^3$ volume. Having defined our mock photometric LAEs and mock spec-z sources, we can now create a 3D reconstruction of a given cosmic structure whose dark matter distribution is known. 

\begin{deluxetable*}{ccccc}
    \tablecaption{Summary of $D_{\rm TV}$ values \label{tab:kl_tvd}}
    \tablehead{
        \colhead{\makecell{Group ID \\ ($z = 0$)\tablenotemark{a}}} & 
       \colhead{\makecell{$M_{\rm tot}^{z=0}$ \\ $[M_\odot]$}} & 
        \colhead{\makecell{$M_\ast^{z=0}$\\ $[M_\odot]$}} & 
       \colhead{\makecell{  $\overline{D}_{\rm TV,spec}$\\}} & 
        \colhead{\makecell{$\overline{D}_{\rm TV,full}$\\ }} 
    }
    \startdata
    0 & $1.54\times 10^{15}$ & $2.89\times 10^{13}$ & 0.245 $\pm$ 0.002 & 0.187 $\pm$ 0.001  \\
    1 & $1.31\times 10^{15}$ & $1.60\times 10^{13}$ & 0.341 $\pm$ 0.003 & 0.300 $\pm$ 0.003  \\
    2 & $1.03\times 10^{15}$ & $1.29\times 10^{13}$ & 0.317 $\pm$ 0.002 & 0.261 $\pm$ 0.002  \\
    3 & $9.00\times 10^{14}$ & $1.26\times 10^{13}$ & 0.271 $\pm$ 0.002 & 0.198 $\pm$ 0.002  \\
    4 & $8.42\times 10^{14}$ & $1.12\times 10^{13}$ & 0.305 $\pm$ 0.002 & 0.232 $\pm$ 0.001  \\
    14 & $4.37\times 10^{14}$ & $6.96\times 10^{12}$ & 0.353 $\pm$ 0.002 & 0.276 $\pm$ 0.001  \\
    \enddata
    \tablenotetext{a}{For six TNG300 protoclusters at $z=3$, total variation distance, $D_{\rm TV}$, is measured  relative to the underlying DM distribution. Reconstruction is conducted using spec-z sources only (`spec') and using the full mock LAE sample (`full').}
    \vspace{-5pt}
\end{deluxetable*}

Figure~\ref{fig:tng300} shows the result for the most massive cosmic structure in TNG300. The top panels show smoothed dark matter density in a cosmic slice with a thickness of 20~cMpc (with the smaller thickness being chosen to better highlight the massive structure and used only for visualization), while panels (b) and (c) show mock spec-z sources and all mock LAEs, respectively. The reconstruction results of the same structure based on the distribution of DM particles, spec-z sources, and all LAEs are shown in panels (d), (e), and (f), respectively. To even out the mismatch of spatial resolution, we smooth the dark matter distribution with a Gaussian kernel with FWHM half of that used for galaxy reconstruction (1--1.5~cMpc). 
The spec-z-only reconstruction, visualized in panel (e), shows noticeable gaps that do not align well with the dark matter structure. This misalignment is expected due to the limited number of spectroscopically confirmed sources. In contrast, the reconstruction utilizing the full LAE sample -- shown in panel (f) -- exhibits a stronger agreement with the dark matter distribution. This is particularly evident in the central regions, where overdensities align more closely with the underlying mass distribution. We show later in this section that the full reconstruction provides a statistically significant improvement over the spec-z reconstruction.

It is also evident in Figure~\ref{fig:tng300} that structures present in the reconstructed map are strongly influenced by the mock LAE and spec-z assignments. 
This means that the final density map can vary not only with the initial conditions but also has features that do not reasonably match the underlying matter distribution. For example, a small LAE overdensity located at $(X,Y)\approx (30,10)$~cMpc seen in panel (c) occurred for no reason other than a random alignment of mock LAEs. Still, the presence of this overdensity led to the assignment of four spec-z sources in the region, which shows up in both spectroscopic and probabilistic reconstruction (panels (e) and (f)). The said feature would not appear if we redo the LAE selection, however. 

To quantitatively evaluate the similarity between the reconstructed distributions and the underlying matter distribution, we use the Total Variation Distance, $D_{\rm TV}$ \citep{Tsybakov_2009}. It quantifies the maximum discrepancy between two probability distributions, ranging from 0 (for identical distributions) to 1 (for disjoint distributions), and for two discrete probability distributions $P_1(\vec{x})$ and $P_2(\vec{x})$ is defined as:
\begin{equation}\label{eq:dtv}
    D_{\rm TV} \equiv \frac{1}{2}\sum_i |P_1(\vec{x_i}) - P_2(\vec{x_i})|
\end{equation}
We choose the Total Variation Distance over other metrics, such as the Kullback-Leibler divergence \citep{Kullback_Leibler1951} or the $\chi^2$-distance, because it is a simple, linear measure of the difference between two distributions.  Logarithmic measures like the Kullback-Leibler divergence are ill-suited to our data, which includes several zero-valued cells and therefore results in undefined or infinite values. Similarly, the $\chi^2$-distance disproportionately emphasizes discrepancies in low-density regions, where the number of galaxies is small, making it less suitable for our analysis.

We calculate $D_{\rm TV}$ for the spec-z-only and full-LAE 3D reconstructions relative to the smoothed dark matter distribution within the (60~cMpc)$^3$ box (with the smoothing scale for dark matter being half of that used for galaxies, as noted above). We repeat the calculations for 100 realizations of the reconstructed distribution to mitigate the biases that may be introduced from random assignments. Table~\ref{tab:kl_tvd} lists the average $D_{\rm TV}$ values from these 100 realizations for the five largest structures ($\log(M_{\rm tot}^{z=0}/M_\odot) \gtrsim 15.0$) and one smaller structure ($\log(M_{\rm tot}^{z=0}/M_\odot) \approx 14.6$) in TNG300 selected by \citet{andrews2024}. 
We provide the mean and standard deviation estimated from all realizations to illustrate that the overall variations in reconstructed distributions are relatively small. We find that the $D_{\rm TV}$ values are uniformly lower when the reconstruction makes use of the full LAE sample compared to spectroscopically confirmed galaxies only, suggesting that the probabilistic reconstruction recovers structural information missed by incomplete and sparse spectroscopic coverage. In Section~\ref{subsec:confirmed_fraction}, we discuss how total variation distance changes with source density ($\bar{\Sigma}_{\rm LAE}$) and the fraction of spectroscopic sources.



\begin{figure*}[t]
    \centering
    \includegraphics[width=0.9\linewidth]{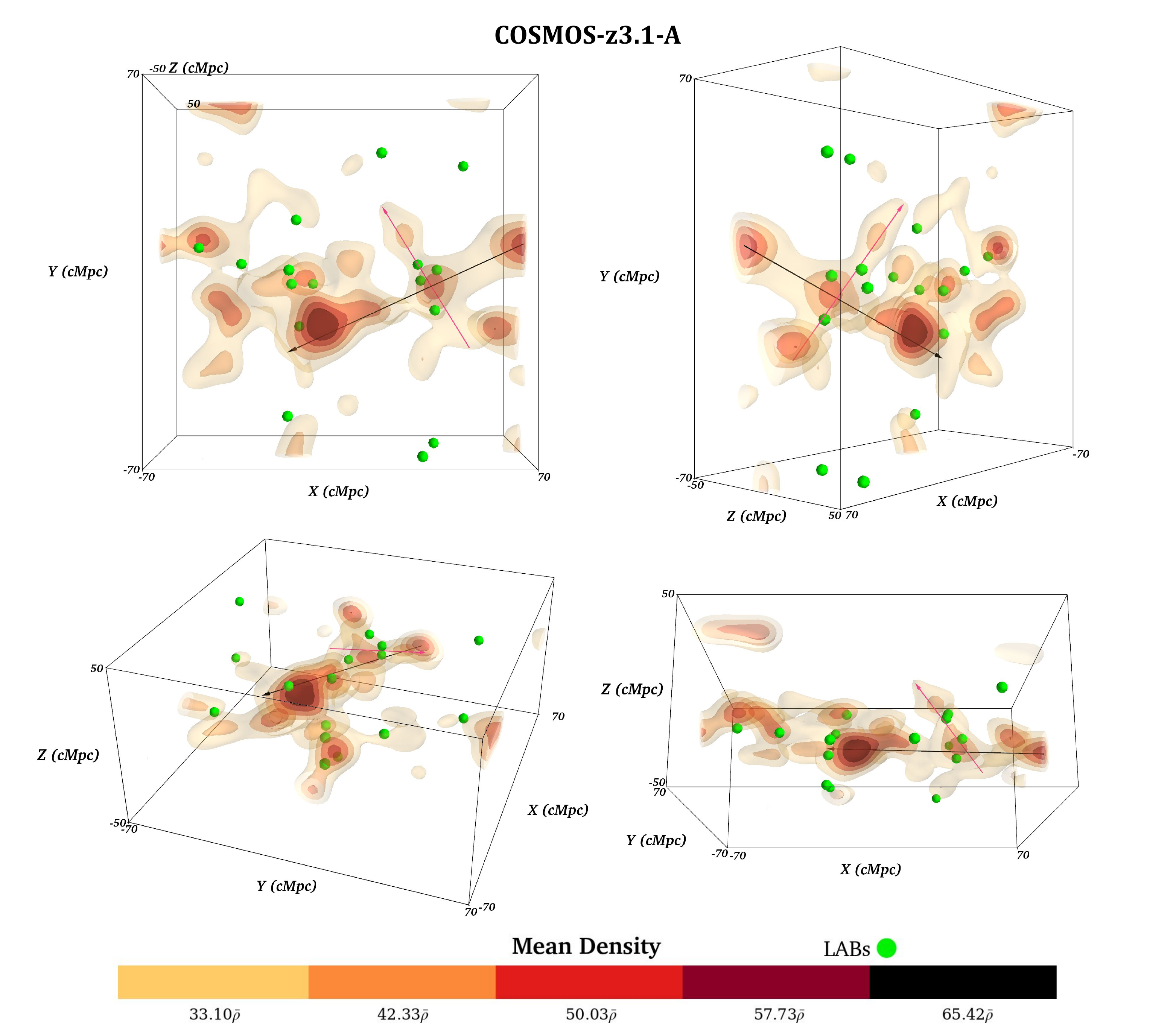}
    \caption{The 3D reconstruction of COSMOS-z3.1-A showing four various angles, plotted in {\tt Mayavi}. The X- and Y-axis represent the east-west and north-south directions respectively (such that a larger X value indicates a position further to the west of the structure, and a larger Y value a position to the north), while the Z-axis represents the line-of-sight direction (such that a larger Z value indicates a higher redshift and correspondingly a greater distance from the observer). The different colored contours represent different volume overdensities relative to the mean volume density $\rho$ across the entire field in the redshift range sampled by $N501$. Green spheres represent spectroscopically confirmed LABs. The black and red arrows trace two filamentary structures for a visual aid.}
    \label{fig:cosmos_n501_s1_3d}
\end{figure*}

\section{ODIN structures in 3D} \label{sec:odin_structures} 


Having established that the probabilistic reconstruction described in Section~\ref{sec:reconstruction} reliably recovers the 3D structures of protoclusters, we apply this technique to the two ODIN clusters and construct their 3D density maps. The visualizations\footnote{Interactive versions of these 3D visualizations are available on GitHub (\href{https://ortiz140.github.io/odin/}{https://ortiz140.github.io/odin/}).}, shown in Figures~\ref{fig:cosmos_n501_s1_3d} and \ref{fig:cosmos_n501_s2_3d}, are made using the software {\tt mayavi} \citep{mayavi}. We also include confirmed LABs in these maps to demonstrate the close physical association between protoclusters, filaments, and LABs as reported in \citet{Ramakrishnan2023}. In this section, we discuss our findings.

\subsection{COSMOS-z3.1-A}\label{subsec:complex_A}

COSMOS-z3.1-A is among the largest cosmic structures identified by ODIN so far. This region contains eight distinct LAE surface density peaks -- shown in the top left panel of Figure~\ref{fig:spec_sources} as white contours -- spanning approximately $45\arcmin\times 45\arcmin$ in the sky, interconnected by cosmic filaments \citep{Ramakrishnan2023}. It hosts around 400 LAEs and 17 LABs, of which we have spectroscopically confirmed 249 LAEs and 13 LABs. Our data firmly establish COSMOS-z3.1-A as an exceptionally rare and highly overdense region—a proto-supercluster. The only comparable system, {\it Hyperion}, is observed at a later epoch \citep[$z\approx 2.4$,][]{Cucciati2018}.

Recently, \citet{McConachie2025} reported the discovery of $MAGAZ3NE$~J100143+023021 {($MAG-1001$ henceforth)}, a protocluster at $z=3.122$, initially identified using the COSMOS2020 photometric redshift catalog \citep{Weaver2022:COSMOS2020} and subsequently confirmed through Keck/MOSFIRE spectroscopy. The spectroscopically confirmed region largely overlaps with one of the eight surface density peaks COSMOS-z3.1-A. While a few other ODIN peaks align with the photo-z density map, they lack spectroscopic confirmation \citep[see Figure~3 of][]{McConachie2025}, and the remaining peaks fall outside the 1.6~deg$^2$ COSMOS field. Among the 28 confirmed galaxies, three are ultramassive ($\log(M_\ast/M_\odot)\gtrsim11$). {As we discuss later in this section, this structure is reasonably recovered in our 3D reconstruction.} The presence of ultramassive galaxies and numerous LABs in COSMOS-z3.1-A suggests intense star-formation and/or AGN activity within these dense protocluster environments.  



Figure~\ref{fig:cosmos_n501_s1_3d} presents a 3D reconstruction of COSMOS-z3.1-A from four different perspectives. The 3D contours are color-coded to reflect spatial density relative to the mean value. The X-Y plane represents the transverse (angular) dimensions, with the X-axis oriented along the east-west direction and the Y-axis along the north-south direction. The Z-axis aligns with the line-of-sight (redshift) direction, where negative values indicate positions closer to us and positive values indicate farther. As a visual aid, two fixed arrows are placed within the volume: the red arrow is aligned with a cosmic filament extending northwest from a large density peak, 
while the black arrow connects three prominent overdensity peaks. 
Green spheres mark the locations of confirmed LABs. Slight opacity has been applied to the contours to emphasize the spatial depth of the structures.



\begin{figure*}
    \centering
    \includegraphics[width=\linewidth]{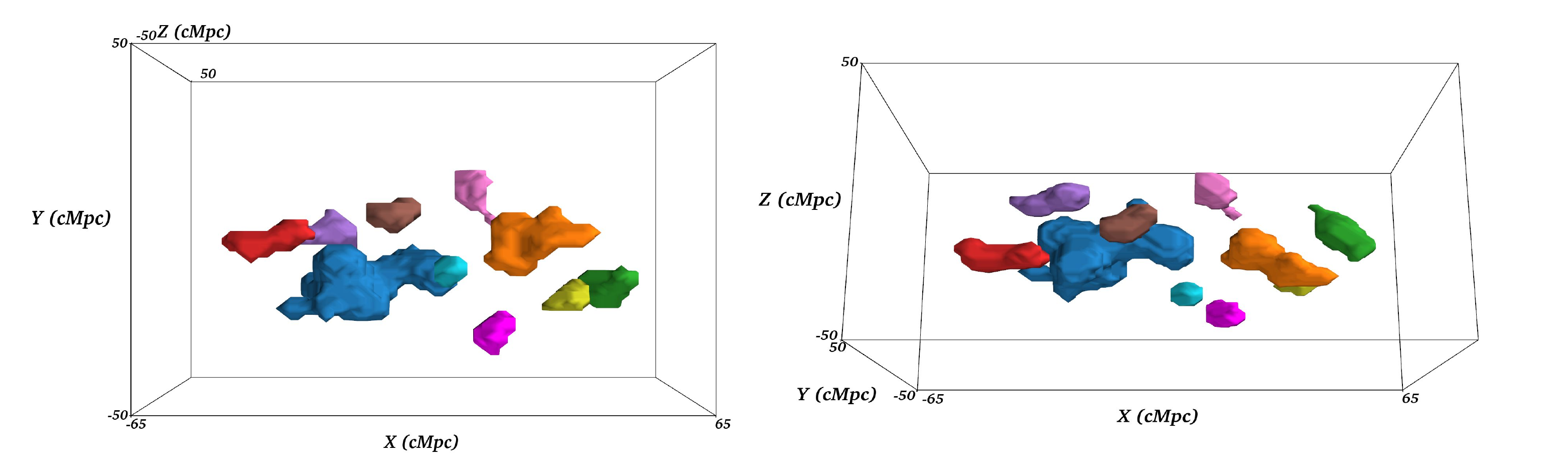}
    \caption{3D perspective of the ten density peaks making up COSMOS-z3.1-A, with each peak given a unique color. The left panel shows the density peaks face-on (as in the top-left panel of Figure \ref{fig:cosmos_n501_s1_3d}) while the right panel shows them as seen along the same line-of-sight as the bottom-right panel of Figure \ref{fig:cosmos_n501_s1_3d} }
    \label{fig:3d_peaks}
\end{figure*}

Based on our 3D reconstructions, we identify ten distinct density peaks (Figure \ref{fig:3d_peaks}). Density peaks are identified based on a spatial density and volume threshold, in a conceptually similar manner to source detection in astronomical images, where an intensity threshold and a minimum isophotal area define detected sources. However, in our case, we ensure the stability and robustness of the identified peaks by performing 100 iterations of the reconstruction and only retaining those voxels which are above our chosen threshold in over 95\% of the reconstructions. In other words, we define each density peak as a contiguous region of voxels with a minimum volume of 500~cMpc$^3$, where the spatial density of each voxel is consistently in the top 3\% of the total volume shown in Figure~\ref{fig:cosmos_n501_s1_3d}. The density threshold is further justified in Section~\ref{subsec:descendant_mass}, while the volume threshold is set to match a sphere with a radius of 5~cMpc—a typical size for protoclusters at $z = 3$ \citep[see, e.g.,][]{Chiang2013,muldrew15}. 

\begin{figure*}
    \centering
    \includegraphics[width=0.8\linewidth]{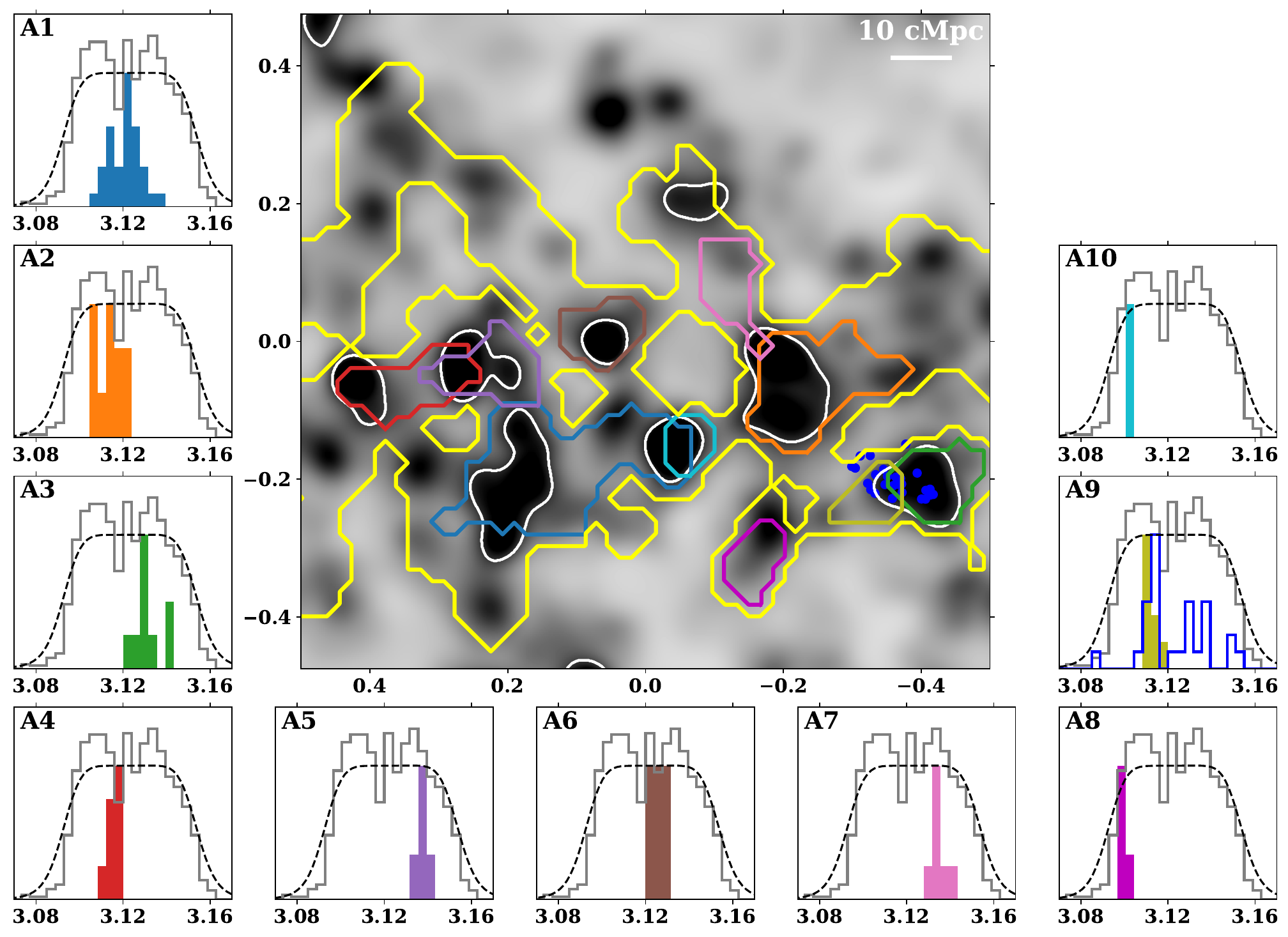}
    \caption{\emph{Main panel:} Contours of the 3D density peaks of COSMOS-z3.1-A in projection (shown in color, with the color used for each peak corresponding to that used in Figure \ref{fig:3d_peaks}). The white contours indicate the locations of the 2D protocluster candidates. The yellow contour outlines the region we consider to be the full volume of the proto-supercluster (see text for details). 
    Dark blue points show member galaxies of the protocluster $MAGAZ3NE$ J100143+023021 spectroscopically confirmed by \citet{McConachie2025}. We recover this structure as peak A9.
    \emph{Side panels:} Redshift distribution of confirmed LAEs within each density peak, shown in the same color as the corresponding contour in the main panel. The grey histogram indicates the redshift distribution of all spectroscopically confirmed LAEs, while the black curve shows the filter transmission of the $N501$ filter. The dark blue histogram in panel A9 shows the redshift distribution of protocluster galaxies confirmed by \citet{McConachie2025}. 
    }
    \label{fig:cosmos_z3.1_a}
\end{figure*}

Figure~\ref{fig:cosmos_z3.1_a} presents the projection of these ten 3D density peaks onto the sky, represented as color contours. The corresponding redshift distributions of spec-z sources within each peak are displayed in smaller panels, using matching colors. {Each peak contains at least three spec-z sources, forming well-defined redshift peaks with widths $\Delta z \lesssim 0.04$, consistent with other spectroscopic studies of protoclusters \citep[e.g.,][]{Dey2016,Toshikawa2016, Cucciati2018}.} Notably, the overall redshift distribution across the entire field is significantly broader than that of the individual peaks and aligns well with the expected transmission of the $N501$ filter. {We also show the spectroscopically confirmed member galaxies of $MAG-1001$ as dark blue dots in the main panel of Figure~\ref{fig:cosmos_z3.1_a}, and plot their redshift distribution in the same color in the side panel labelled `A9'. Of the 28 galaxies confirmed by \cite{McConachie2025}, nine fall within the density peak we designate A9, forming a clear spike in redshift space as shown in the side panel of Figure~\ref{fig:cosmos_z3.1_a}. Notably, the most massive galaxy within $MAG-1001$, with stellar mass of log$(M/M_\odot) \sim 11.15$, is among these nine sources. The remaining galaxies are more broadly distributed in redshift; five fall within the density peak A3, while an additional seven fall at line-of-sight positions between A3 and A9, aligned with a moderate density bridge connecting the two density peaks. The remaining galaxies are in the foreground or background relative to the overdensity we detect.}

From our reconstruction, we estimate the descendant mass of each density peak using the following relation \citep{Steidel1998,Steidel2000}:
\begin{equation}\label{eq:descendant_mass}
    M^{z=0}_{\rm est} = \left(1+\frac{\delta_{g}}{b_{g}}\right) \rho_{0}V_{PC}
\end{equation}
where $V_{PC}$ is the protocluster volume in comoving units (in this case the volume with density in the top 3\%) and $\rho_{0}$ represents the mean comoving matter density of the Universe. $\delta_g$ is the three-dimensional galaxy overdensity and is related to matter overdensity ($\delta_m$) as $\delta_g = b_g \delta_m$.  We fix the galaxy bias value to $b_g=1.8$, consistent with clustering measurements of ODIN $N501$ LAEs \citep{White2024,herrera25}. By leveraging cosmic structures in TNG300, we estimate an uncertainty of $\approx$0.2--0.3~dex in the resultant values of $M^{z=0}_{\rm est}$ (see Section~\ref{subsec:descendant_mass}). {We find that raising (lowering) the density threshold used to define the peaks to the top 2\% (5\%) changes the estimated descendant masses by less than this uncertainty, with the mass decreasing (increasing) by $\sim$ 0.10~dex (0.15~dex). We hence conclude that our descendant mass estimates are reasonably robust against the specific choice of density threshold, and the error budget is dominated by the uncertainty in protocluster membership.} 

The centroid of each peak is determined as the density-weighted mean of its constituent voxels, with the associated uncertainty derived from its second moment. These properties are detailed in Table~\ref{tab:density_peaks}. {The cumulative mass of the ten density peaks of COSMOS-z3.1-A is approximately $\log(M/M_\odot) = 15.5$ (see Table \ref{tab:density_peaks})}, twice the mass of Coma \citep[e.g.,][]{Ho2022,Kang2025} and $\sim 1.5$ times the mass of {\it El Gordo} \citep[e.g.,][]{Kim2021}, a well-known merging cluster system at $z=0.87$. This remarkable concentration of mass strongly suggests that COSMOS-z3.1-A is the progenitor of an exceptionally rare and ultra-massive cosmic structure.

\begin{deluxetable*}{cccccccc}[t]
    \tablecaption{{Density peaks in COSMOS-z3.1-A and COSMOS-z3.1-C}\label{tab:density_peaks}}
    \tablehead{\colhead{Structure ID} & \colhead{R.A.\tablenotemark{a}} & \colhead{Decl.\tablenotemark{a}} & \colhead{$z_{\rm peak}$\tablenotemark{a}} &\colhead{$N_{\rm spec}$} & \colhead{$V_{PC}$ (10$^3$ cMpc$^3$)\tablenotemark{b}} & \colhead{$\delta_g$\tablenotemark{c}} & \colhead{$\log {M^{z=0}_{\rm est,3D}}/{M_\odot}$\tablenotemark{d}}}
    \startdata
    \hline
    \multicolumn{8}{c}{\textbf{COSMOS-z3.1-A}} \\
    \hline
    COSMOS-z3.1-A1 & 150.932 $\pm$ 0.080 & 2.537 $\pm$ 0.044 & 3.121 $\pm$ 0.006 & 34 & 8.48 & 14.7 & 15.1 \\
    COSMOS-z3.1-A2 & 150.558 $\pm$ 0.51 & 2.663 $\pm$ 0.043 & 3.115 $\pm$ 0.005 & 11 & 3.18 & 13.1 & 14.7 \\
    COSMOS-z3.1-A3 & 150.376 $\pm$ 0.030 & 2.520 $\pm$ 0.028 & 3.130 $\pm$ 0.005 & 9 & 2.18 & 14.6 & 14.5  \\
    COSMOS-z3.1-A4 & 151.144 $\pm$ 0.053 & 2.664 $\pm$ 0.024 & 3.116 $\pm$ 0.003 & 8 & 1.79 & 13.5 & 14.4  \\
    COSMOS-z3.1-A5 & 151.020 $\pm$ 0.041 & 2.682 $\pm$ 0.025 & 3.138 $\pm$ 0.003 & 10 & 1.55 & 14.2 & 14.4  \\
    COSMOS-z3.1-A6 & 150.866 $\pm$ 0.031 & 2.739 $\pm$ 0.023 & 3.126 $\pm$ 0.003 & 3 & 1.33 & 13.9 & 14.3 \\
    COSMOS-z3.1-A7 & 150.679 $\pm$ 0.024 & 2.817 $\pm$ 0.038 & 3.135 $\pm$ 0.003 & 7 & 1.33 & 13.6 & 14.3 \\
    COSMOS-z3.1-A8 & 150.644 $\pm$ 0.022 & 2.404 $\pm$ 0.028 & 3.099 $\pm$ 0.003 & 4 & 1.03 & 13.7 & 14.2 \\
    COSMOS-z3.1-A9 & 150.471 $\pm$ 0.024 & 2.497 $\pm$ 0.022 & 3.111 $\pm$ 0.003 & 8 & 0.73 & 12.9 & 14.0 \\
    COSMOS-z3.1-A10 & 150.736 $\pm$ 0.020 & 2.577 $\pm$ 0.020 & 3.103 $\pm$ 0.002 & 4 & 0.64 & 14.1 & 14.0 \\
    \hline
    \textbf{Total} & -- & -- & -- & \textbf{98} & \textbf{22.24} & -- & \textbf{15.5} \\
    \hline\hline
    \multicolumn{8}{c}{\textbf{COSMOS-z3.1-C}} \\
    \hline
    COSMOS-z3.1-C1 & 148.723 $\pm$ 0.093 & 2.731 $\pm$ 0.064 & 3.116 $\pm$ 0.014 & 33 & 9.66 & 14.8 & 15.2 \\
    COSMOS-z3.1-C2 & 148.881 $\pm$ 0.035 & 2.420 $\pm$ 0.026 & 3.118 $\pm$ 0.004 & 6 & 2.21 & 15.9 & 14.5 \\
    COSMOS-z3.1-C3 & 149.694 $\pm$ 0.057 & 2.481 $\pm$ 0.025 & 3.104 $\pm$ 0.002 & 5 & 1.79 & 14.3 & 14.4 \\
    COSMOS-z3.1-C4 & 148.907 $\pm$ 0.035 & 2.328 $\pm$ 0.021 & 3.103 $\pm$ 0.002 & 5 & 1.00 & 13.6 & 14.1 \\
    \hline
    \textbf{Total} & -- & -- & -- & \textbf{50} & \textbf{14.66} & -- & \textbf{15.3} \\
    \enddata
    \tablenotetext{a}{The centroid of each peak estimated from the 3D reconstruction. The uncertainties are computed as the density-weighted second moment of the positions of the voxels.}
    \tablenotetext{b}{The volume of the density peak above a density threshold of 97\%, see text.}
    \tablenotetext{c}{The median 3D galaxy overdensity within the density peak. The galaxy overdensity is overestimated due to the effect of peculiar motion (see Section~\ref{subsec:descendant_mass} and Appendix B).}
    \tablenotetext{d}{The descendant mass is corrected for overestimation by a factor of two. The uncertainty in the estimated descendant mass is $\sim$0.2--0.3~dex (see Section~\ref{subsec:descendant_mass}).}
\end{deluxetable*}

{In addition to the density peaks, the 3D reconstruction also displays lower-density features connecting the peaks together. As seen in the bottom-right panel of Figure \ref{fig:tng300}, these features may not be as well represented by the 3D reconstruction as the most overdense peaks, and confirming that they are truly coherent structures would require additional spectroscopy. However, the features of the 3D reconstruction which correspond to filaments detected from the projected positions of LAEs (shown in Figure \ref{fig:spec_sources}) are more likely to represent true filaments of the cosmic web. In \citet{Ramakrishnan2024}, we showed that the 2D filamentary network traced by photometrically selected LAEs is a reliable representation of the true 3D cosmic web; further, we found that 2D filaments in close projected proximity to a density peak are extremely likely (with a probability of $\gtrsim 95\%$) to be connected to that peak. It is thus possible that several of the filamentary structures seen in Figure \ref{fig:cosmos_n501_s1_3d}, particularly the two highlighted by the black and red arrows, are true features of the 3D LSS.}

As stated previously, COSMOS-z3.1-A is the largest LAE overdensity identified by ODIN to date. The total cosmic volume surveyed by ODIN at present is $\sim$ 2.2 $\times$ 10$^7$ cMpc$^3$ \citep[comprising 4 field/filter combinations, see][]{Ramakrishnan2025}; this restricts the volume density of such structures to be $\lesssim 5 \times 10^{-8}$~cMpc$^{-3}$, that is, rarer than even the most massive cluster progenitors. 
Such extreme structures serve as an important test for cosmological models, as they constrain the magnitude of large-scale density fluctuations \citep[e.g.,][]{Park2012,Hwang2016}.
The number and size of the largest overdensities identified by ODIN may thus provide another avenue to constrain cosmological parameters.


\subsection{COSMOS-z3.1-C}

\begin{figure*}
    \centering
    \includegraphics[width=0.9\linewidth]{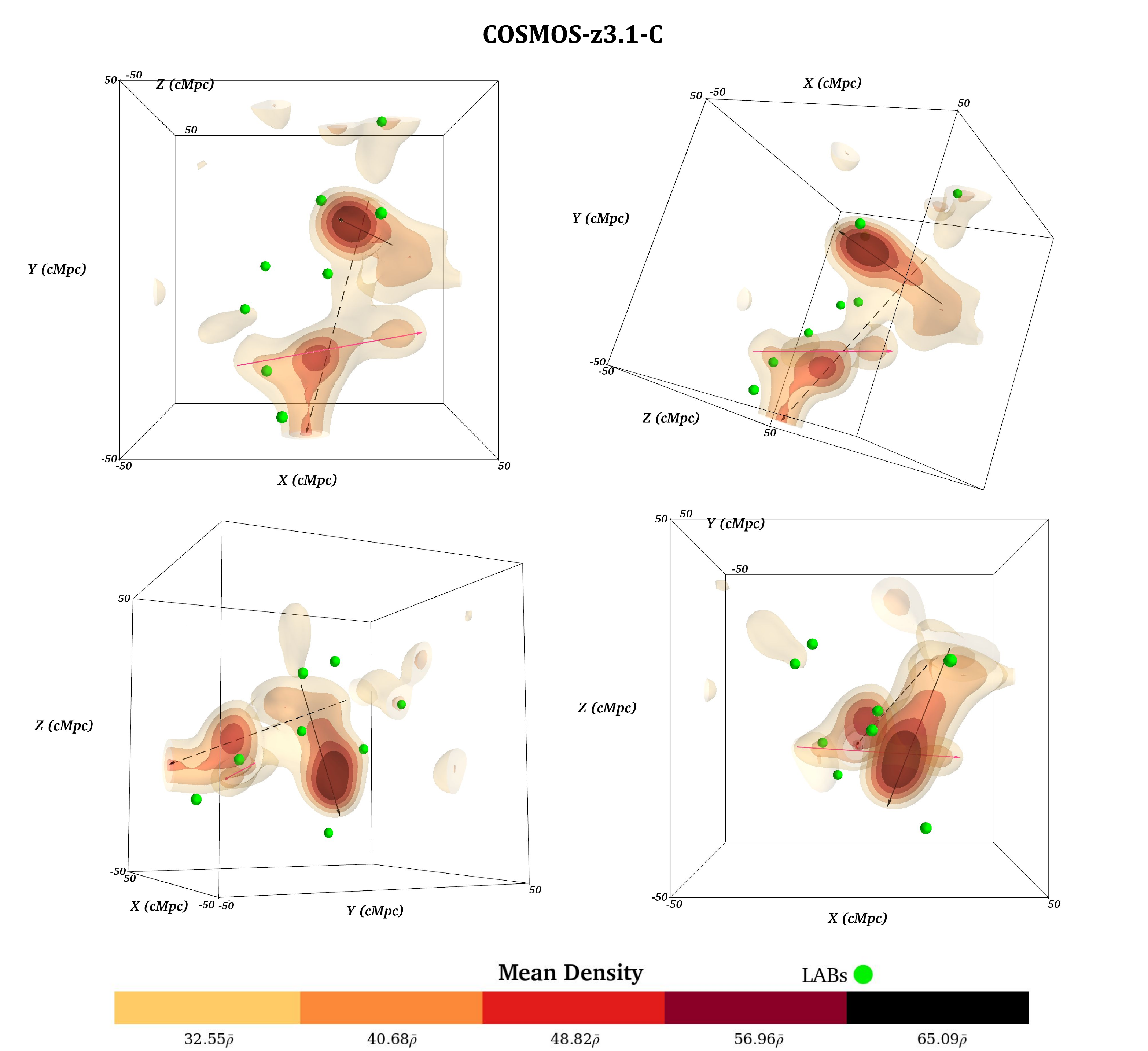} 
    \caption{As Figure \ref{fig:cosmos_n501_s1_3d}, but of COSMOS-z3.1-C. The black, red, and dashed arrows are added for visual aid. The top left panel is aligned with the observed sightline. }
    \label{fig:cosmos_n501_s2_3d}
\end{figure*}

The surface density map of COSMOS-z3.1-C, shown in the bottom panels of Figure~\ref{fig:spec_sources}, reveals a distinctive morphology characterized by a prominent high-density region connected to an almost linear structure. A similar configuration was recently observed in a structure at $z=3.44$, dubbed {\it Cosmic Vine} \citep{Jin2024}. However, the three-dimensional structure tells a more complex story (see Figure~\ref{fig:cosmos_n501_s2_3d}). The main density peak, which appears compact and round in projection, actually consists of a dense core at the `front' (i.e., closer to us) with a tail extending along the direction indicated by the solid black arrow in Figure~\ref{fig:cosmos_n501_s2_3d}. Below this are three closely-connected density peaks stretching along the direction indicated by a red arrow in Figure~\ref{fig:cosmos_n501_s2_3d}. The northern and southern portions of the structure also appear to be linked toward the rear of the observed volume. {As in the case of COSMOS-z3.1-A, this lower-density bridge might be a true feature of the LSS given its proximity to the density peaks and its correspondence to a 2D filament traced by photometric LAEs.}

\begin{figure*}
    \centering
    \includegraphics[width=0.65\linewidth]{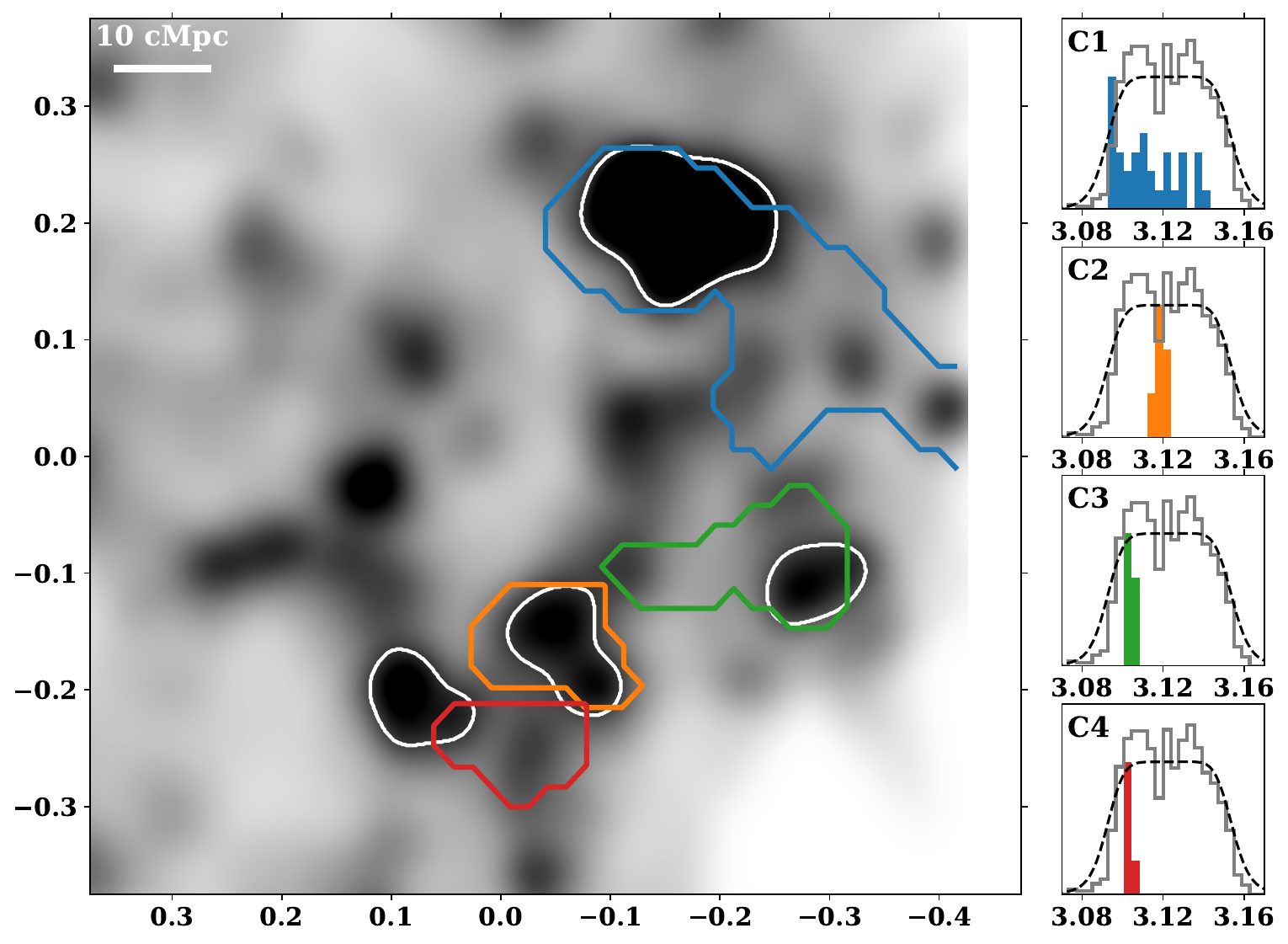}
    \caption{As Figure \ref{fig:cosmos_z3.1_a}, but for COSMOS-z3.1-C}
    \label{fig:cosmos_z3.1_c}
\end{figure*}

The four peaks identified in 3D are projected onto the sky and displayed in Figure~\ref{fig:cosmos_z3.1_c} as colored contours, along with their corresponding redshift distributions. Their coordinates and estimated descendant masses are provided in Table~\ref{tab:density_peaks}. The combined mass of COSMOS-z3.1-C within these overdensities is $\log(M/M_\odot), \approx 15.3$, $\sim$1.5 times lower than that of COSMOS-z3.1-A.


The structure of COSMOS-z3.1-C offers a compelling example of how the detection and interpretation of cosmic structures are strongly influenced by the observer’s line of sight. It also underscores the irregular and clumpy nature of large-scale structure formation at high redshift. From our viewing angle, the primary peak C1 appears as a compact, highly overdense protocluster. However, as shown in the top-right panel of Figure~\ref{fig:cosmos_n501_s2_3d}, viewing it from a different perspective would reveal a more elongated and diffuse structure, similar in size to the combined extent of the three smaller peaks. Similarly, the bottom panels of Figure~\ref{fig:cosmos_n501_s2_3d} suggest that peaks C2, C3, and C4 — while extended and distinguishable as individual peaks in the current view — could appear as a single, compact system with a comparable descendant mass to C1 when observed from an alternate angle. 


\section{Discussion} \label{sec:discussion}

\subsection{COSMOS-z3.1-A: A supercluster in formation}


In this section, we discuss how COSMOS-z3.1-A compares to {\it Hyperion} \citep{Cucciati2018}, the only other known high-redshift proto-supercluster. Existing data firmly established that both contain several density peaks, seven in {\it Hyperion} and ten in COSMOS-z3.1-A. However, it is not straightforward to compare the masses of these individual peaks. Although all {\it Hyperion} peaks, like COSMOS-z3.1-A, have spec-z members, 
the majority of the galaxies used to trace this structure only have photometric redshifts. The uncertainty in the redshift of individual sources is $\Delta z\approx 0.1-0.2$, i.e., several times larger than what the ODIN filter transmission allows ($\Delta z\approx 0.06$). The larger redshift uncertainty effectively requires a higher density threshold to robustly identify peaks and thus leads to smaller structure volumes and smaller total masses for individual peaks. 
Indeed, the mass estimates of {\it Hyperion}'s peaks range in mass from $\log(M/M_\odot) \sim 13.0 - 14.4$ \citep[see Table~2 of][]{Cucciati2018}, much smaller than the $\log(M/M_\odot) \sim 14.0 - 15.1$ listed in Table~\ref{tab:density_peaks}. These differences are challenging for us to properly model and correct for, to enable a fairer comparison. 

{\citet{Cucciati2018} estimate the total mass of {\it Hyperion} to be $\log(M/M_\odot) \sim 15.7$ by considering the volume enclosed above a density threshold of $2\sigma$. By comparison, the total mass within the ten density peaks of COSMOS-z3.1-A is $\log(M/M_\odot) \sim 15.5$. The total mass above a density threshold of 92\% --chosen to delineate a volume containing all ten density peaks, and shown by the yellow contour in Figure \ref{fig:cosmos_z3.1_a} -- is $\log(M/M_\odot) \sim 16.0$. Thus, we estimate the total mass of COSMOS-z3.1-A to be in the range $\log(M/M_\odot) \sim 15.5 - 16.0$, comparable to {\it Hyperion}. Likewise, the total volume above the 92\% density threshold is $\sim 9.8 \times 10^4$~cMpc$^3$, similar to the $9.5 \times 10^4$~cMpc$^3$ volume of {\it Hyperion}.} 

{In order to ensure that this extremely large total mass estimate is not due to artifacts introduced by our reconstruction process, we consider a similar volume to COSMOS-z3.1-A (100$\times$100$\times$60~cMpc$^3$) around the $z = 3$ progenitors of the 30 most massive clusters in TNG300. As described in Section \ref{subsec:validation}, we designate a fraction of the mock LAEs in this volume to be spec-z LAEs while the remainder are considered to be photometric LAEs. Additionally, in consideration of the fact that some fraction of the unconfirmed ODIN LAEs may in fact be interlopers at a different redshift, we replace a number of the mock photometric LAE sample equal to 5\% of the total number of mock LAEs with randomly distributed `interlopers'. We then carry out the reconstruction process and estimate the total mass above a density threshold of 92\%, as for COSMOS-z3.1-A. We find that in all cases, the mass contained above this threshold is $\log(M/M_\odot) \lesssim 15.5$. This suggests that COSMOS-z3.1-A is legitimately an extremely massive structure.}

\begin{figure*}
    \centering
    \includegraphics[width=1.0\linewidth]{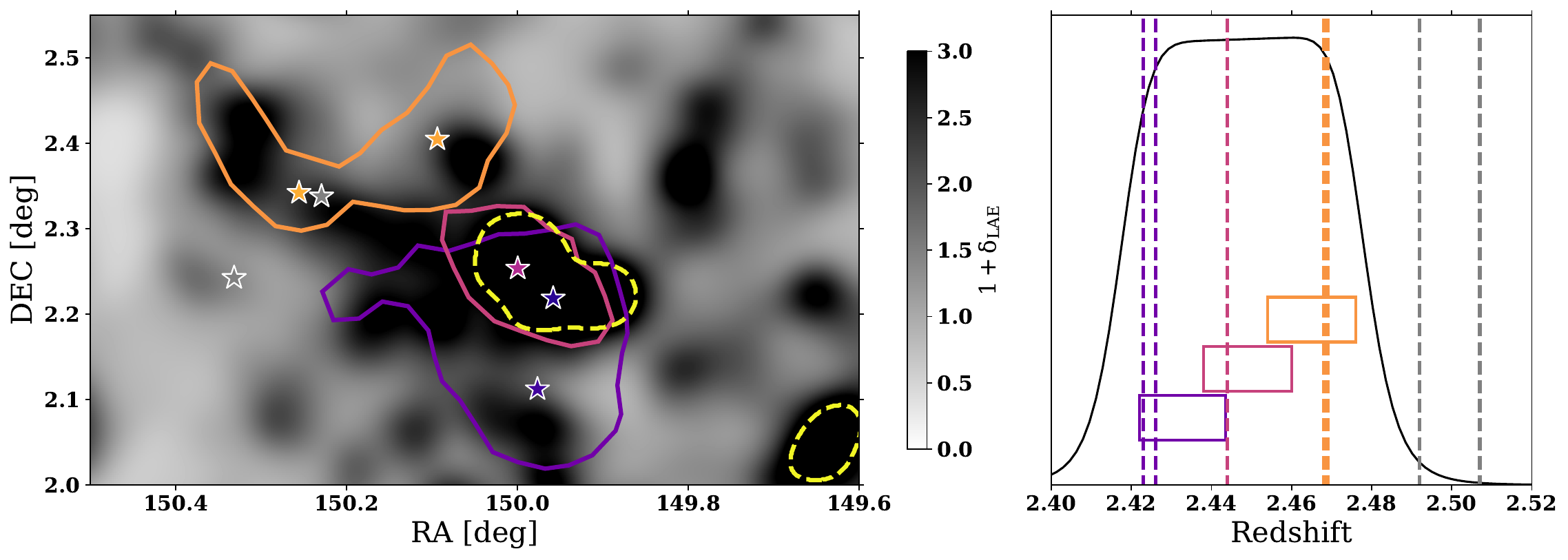}
    \caption{The surface density map of ODIN $N419$ LAEs in the field of {\it Hyperion}. The protocluster detected by ODIN is highlighted by a yellow contour, while the 2D overdensities detected by \protect{\citet{Cucciati2018}} (other colored contours) and the positions of the individual density peaks (colored stars) are shown in comparison. 
    {Rectangular blocks} in the right panel indicate the redshift ranges of the overdensities indicated by the corresponding colored contour in the left panel. Vertical lines indicate the redshifts of the seven density peaks, {with colors matched to the stars shown in the left panel}. Because {\it Hyperion} is more extended along the line-of-sight direction than our narrowband filter, it does not appear as a significant structure in the ODIN data. 
    }
    \label{fig:hyperion}
\end{figure*}

How likely is it that ODIN detects structures like {\it Hyperion}? Similarly, can redshift surveys identify systems such as COSMOS-z3.1-A? These questions can be explored by examining {\it Hyperion}, which has been studied using both spectroscopy and imaging. The latter includes observations using another ODIN filter, $N419$. {While the lack of publicly available spectroscopic data covering {\it Hyperion} prevents us from applying our 3D reconstruction method to this structure, we can compare its projected extent to the distribution of (photometrically selected) $N419$ LAEs.} The left panel of Figure~\ref{fig:hyperion} shows a surface density map of $N419$-selected LAEs in greyscale. Overlaid are three color contours, adapted from Figure~1 of \citet{Cucciati2018}, which trace the angular extent of {\it Hyperion} across three slightly overlapping redshift slices that span the $N419$ filter transmission range (see right panel). These contours closely align with regions of elevated LAE overdensity. The pink, tracing the middle redshift slice -- corresponding to the peak $N419$ transmission -- closely matches the highest LAE surface density peak, identified as an ODIN protocluster \citep[thick yellow dashed line; see][for details]{Ramakrishnan2024}.

From Figure~\ref{fig:hyperion}, we conclude that the $N419$ data reasonably recover the angular extent of {\it Hyperion}. {This is consistent with the findings of \citet{huang20}, who used a wider narrowband filter tuned to span the full redshift range of {\it Hyperion} and succesfully identify LAE overdensities corresponding to all seven redshift peaks, suggesting that the large-scale structure traced by LAEs and spectroscopically confirmed galaxies is broadly consistent. They noted that this consistency may partly arise from the higher spectroscopic success rate for galaxies with strong Ly$\alpha$ emission}. 

{However, 
unlike COSMOS-z3.1-A, {\it Hyperion} does not display a concentration of multiple overdensities in the surface density map of ODIN LAEs; our fiducial protocluster selection \citep[described in][]{Ramakrishnan2025} identifies only a single object within the region, indicated by a dashed yellow contour in the left panel of Figure \ref{fig:hyperion}. This is not surprising -- as shown by the right panel of Figure \ref{fig:hyperion}, only one of {\it Hyperion's} seven redshift peaks falls squarely within the $N419$ filter transmission; the remaining peaks are either at the edges of the redshift range sampled by the filter, where the transmission is lower, or outside the filter window altogether. }

{{\it Hyperion} spans  $\sim 60 \times 60 \times 150$~cMpc$^3$, with its longest dimension aligned along the line of sight, whereas the FWHM of the $N419$ filter corresponds to a depth of $\Delta z\approx 0.06$ or 75~cMpc. In contrast, COSMOS-z3.1-A, with dimensions of $\sim 90 \times 70 \times 60$~cMpc$^3$, is more extended in the transverse direction (though we note that the 60~cMpc thickness of COSMOS-z3.1-A is entirely determined by the $N501$ filter transmission). As a result, {\it Hyperion} appears as a more compact structure in our surface density maps than COSMOS-z3.1-A. This is an intrinsic limitation of narrowband imaging surveys, which sample thin cosmic slices.}


{As can be seen in Figure~\ref{fig:cosmos_n501_s1_3d} (top left), COSMOS-z3.1-A is most extended in the direction indicated by the black arrow. If our sightline had been different and more closely aligned to the black arrow, similar to that shown in the bottom left panel, COSMOS-z3.1-A would likely have been detected in our LAE surface density maps with an extent comparable to that of {\it Hyperion}. Likewise, the transverse extent of COSMOS-z3.1-A (90~cMpc or 45\arcmin\ at $z=3.12$) is nearly as wide as the area covered by VUDS \citep{lefevre15} and C3V0 \citep{Hung2025}, the deep spectroscopic surveys used to trace {\it Hyperion}, and thus is unlikely to be fully captured in such surveys. These considerations highlight the impact of the sightline on identifying the largest cosmic structures, and demonstrate the limitations and complementary nature of the two methods.}


\subsection{Comparison of ODIN structures with simulations}

How do COSMOS-z3.1-A and C compare to structures in the TNG300 simulation? COSMOS-z3.1-C consists of four density peaks, comprising a total mass of $\log(M/M_\odot) = 15.3$. The dominant peak is C1, with a mass of $\log(M/M_\odot) = 15.2$. Among the remaining three peaks, C2 and C3 have similar masses, while C4 is both the southernmost and the least massive. C1 is separated from the remaining peaks by $\sim 30-50$~cMpc while C2-C4 are separated from each other by 15--30~cMpc. 

There are no directly comparable systems to this complex in TNG300. The most massive $z = 0$ cluster in the TNG300 box is Group 0, with $\log(M_{200}/M_\odot) = 15.2$, similar to the estimated descendant mass of C1. This cluster does not have any neighbors of comparable mass to C2-C4. However, if we consider peaks C2-C4 alone, we can find close simulated analogs. For example, the $z = 3$ progenitors of Groups 7, 11 and 13 \citep[where the group numbers are those at $z = 0$, see][]{andrews2024} are separated by $\sim$ 15 - 30 cMpc. These groups encompass a total mass of $\log(M/M_\odot) = 15.2$, very similar to the observed peaks. At $z = 0$, all three groups draw closer to each other, now being separated by $\sim$10 - 15 cMpc. 



There are no comparable structures to COSMOS-z3.1-A in TNG300. This is unsurprising; as stated in Section \ref{subsec:complex_A}, the number density of structures such as COSMOS-z3.1-A is limited to be $\lesssim 5 \times 10^{-8}$~cMpc$^{-3}$ based on the ODIN data obtained to date. The number of such structures within the (302 cMpc)$^3$ volume of TNG300 is thus $\lesssim 1.4$. Finding simulated analogs to COSMOS-z3.1-A will likely require a much larger simulation box, such as that of FLAMINGO \citep[with box sizes of 1 Gpc and 2.8 Gpc to a side;][]{Schaye2023,Kugel2023}, or zoom-in simulations such as TNG-Cluster \citep[where clusters are selected from a box with 1 Gpc to a side and resimulated;][]{Nelson2024}. This will be the subject of future work.

\subsection{The relation of LABs with the LSS}

Both of the structures presented in this work exhibit a high concentration of LABs in their vicinity, as shown in Figures~\ref{fig:cosmos_n501_s1_3d} and \ref{fig:cosmos_n501_s2_3d}. These figures display only the spectroscopically confirmed LABs. The number density of photometrically selected LAB candidates within COSMOS-z3.1-A and COSMOS-z3.1-C is similar to that seen in other protocluster regions and is $\sim$4 times higher than the average across the field \citep{Moon2026}. While previous studies have reported the strong association of LABs with protoclusters \citep[e.g.,][]{matsuda05,Badescu2017,Ramakrishnan2023,zhang25}, our analysis offers a more comprehensive perspective by leveraging 3D spatial information. 

Figures~\ref{fig:cosmos_n501_s1_3d} and \ref{fig:cosmos_n501_s2_3d} clearly demonstrate that LABs tend to inhabit the outskirts of the density peaks. Even those that appear projected onto regions of highest surface density are revealed by spectroscopy to lie either in front of or behind the 3D overdensities. Quantitatively, the median separation of the spectroscopically confirmed LABs from the center of the nearest density peak is $\sim$17~cMpc (discounting 6 LABs in COSMOS-z3.1-A and 2 LABs in COSMOS-z3.1-C, which are far from the bulk of the overdense region). By comparison, the characteristic size of the density peaks, estimated as $V_{PC}^{1/3}$, is $\sim$10--20~cMpc. Likewise, the median (3D) LAE overdensity at the locations of LABs is $\sim 11.2$, whereas the density threshold used to define these peaks is $\sim 11.4$, and the median overdensity across the 14 density peaks identified across the two complexes ranges from $\sim 12.9-15.9$ (see Table \ref{tab:density_peaks}). 
Thus, although the current LAB sample is insufficient in size to demonstrate any statistical trends, there is strong evidence that LABs are associated with protoclusters, but do not occupy their cores.

The physical origin of this spatial distribution remains uncertain, with proposed explanations including cold accretion onto massive halos \citep[e.g.,][]{Daddi2021,daddi22}, the emergence of proto-group media \citep{Badescu2017}, and AGN triggering via galaxy mergers \citep[e.g.,][]{huang22}. In future work, we will further investigate these connections using expanded samples of LABs and protoclusters.

\section{Guiding Future Spectroscopic Efforts with Simulations}\label{sec:outlook}

In this study, we show that extensive spectroscopic follow-up of two large-scale structures reveals intricate details such as dense cores as well as the morphology and connectivity of adjacent density peaks. The two structures presented here are among tens of similarly extended systems and hundreds of individual protoclusters uncovered by ODIN. Despite the advent of powerful wide-field spectrographs such as Subaru's Prime Focus Spectrograph \citep{Greene2022}, DESI, and the upcoming Spec-S5 \citep{Besuner2025}, it remains impractical to achieve comparably dense spectroscopic sampling for all identified structures. This constraint raises a key question: how should limited observational resources be allocated to balance the scientific value of densely sampling individual structures versus more broadly sampling a larger number of systems? With the primary goal of robustly characterizing LSS and estimating descendant halo masses, we turn to the TNG300 simulations to address this question and inform future spectroscopic survey strategies.

\subsection{Spectroscopic Sampling Density}\label{subsec:confirmed_fraction}

\begin{figure*}
    \centering
    \includegraphics[width=0.95\linewidth]{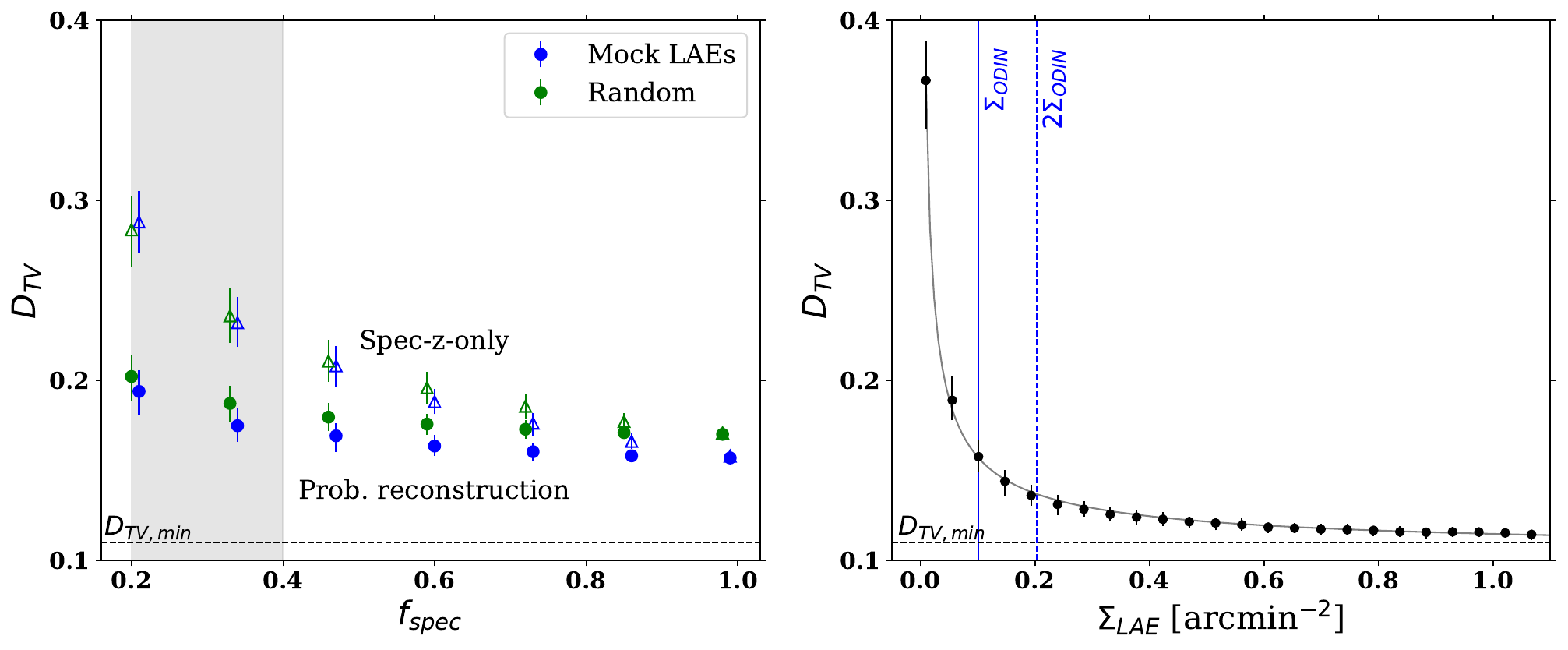} 
    \caption{
{\it Left:} Total variation distance, $D_{\rm TV}$, quantifying the agreement between the 3D galaxy-traced distribution and the underlying dark matter distribution, as a function of the fraction of spectroscopically confirmed sources, $f_{\rm spec}$. The gray shaded region indicates the range of $f_{\rm spec}$ in COSMOS-z3.1-A and C. The LAE surface density is fixed to the observed value. Two galaxy samples are tested: mock LAEs (blue), selected using stellar-mass-weighted probabilities (see Section~\ref{subsec:validation}), and random samples (green), drawn from all galaxies with $M_\ast \geq 10^7 M_\odot$. Open symbols indicate $D_{\rm TV}$ values for reconstructions using only spectroscopic redshifts. In all cases, the probabilistic reconstruction outperforms the spec-z-only approach.
{\it Right:} $D_{\rm TV}$ values as a function of LAE surface density. The observed ODIN value is indicated by the vertical line. 
}
    \label{fig:conf_frac}
\end{figure*}

To evaluate how accurately LSS can be reconstructed as a function of spectroscopic sampling density, we repeat the analysis described in Section~\ref{subsec:validation}, this time varying $f_{\rm spec}$, the fraction of mock LAEs designated as spectroscopic sources. These `confirmed' galaxies serve as inputs for the 3D probabilistic reconstruction. We expect that increasing $f_{\rm spec}$ should yield reconstructions that more closely resemble the true underlying distribution. We vary $f_{\rm spec}$ between 0.2 and 1.0, where $f_{\rm spec}=1.0$ corresponds to all identified LAEs receiving spectroscopic confirmation. For each value of $f_{\rm spec}$, we randomly select a subset of galaxies as confirmed, perform the reconstruction, and compute the total variation distance, $D_{\rm TV}$, between the reconstructed and true dark matter distributions using Equation~\ref{eq:dtv}. This process is repeated over 500 independent realizations, and the resulting $D_{\rm TV}$ values are averaged.

In addition, we define an alternative subset of galaxies within the protocluster volume. Unlike the mock LAE sample, which includes a stellar-mass-dependent weighting as described in Section~\ref{subsec:validation}, this `random' sample is randomly drawn from all galaxies with $M_\ast \geq 10^7 M_\odot$, chosen such that the mean source density matches that of the LAE sample. The same reconstruction and evaluation procedure is then applied to this sample.

The results are presented in the left panel of Figure~\ref{fig:conf_frac}. As expected, agreement with the true matter distribution improves with increasing values of $f_{\rm spec}$. Both the mock LAE and random galaxy samples trace the large-scale structure similarly, as indicated by their closely matching $D_{\rm TV}$ values. Our findings suggest that even at a relatively low spectroscopic fraction of 20\%, the probabilistic reconstruction significantly outperforms the distribution derived from spectroscopic redshifts alone. Figure~\ref{fig:conf_frac} further shows that the $D_{TV}$ value for the probabilistic reconstruction is surprisingly similar for all values of $f_{\rm spec}$, reducing by $\lesssim 25\%$ on increasing $f_{\rm spec}$ from 0.2 to 1. As $f_{\rm spec}$ approaches unity, the $D_{\rm TV}$ values for the two methods converge, which is expected since fewer galaxies require probabilistic redshift assignments. Ultimately, the accuracy achievable by the reconstruction is constrained by the surface density of the ODIN survey; increasing the number of galaxy tracers will improve the recovery of the LSS (see below).



In the right panel of Figure~\ref{fig:conf_frac}, we investigate the extent to which increasing the source density—by observing to greater depth—can improve the reconstruction accuracy. We repeat the same procedure as before, but increase the number of galaxies assigned as photometric LAEs by up to an order of magnitude relative to the observed value. For simplicity, we assume $f_{\rm spec}=1.0$, i.e., all photometric LAEs receive spectroscopic confirmation. The resulting $D_{\rm TV}$ value initially decreases sharply but asymptotes to a minimum value, $D_{\rm TV, min}$, approximately fit by a function, $D_{\rm TV} = (\Sigma_{\rm LAE}/0.0115)^{-0.674}~+~0.103$ where $\Sigma_{\rm LAE}$ is the LAE surface density in units of arcmin$^{-2}$. {Reducing the voxel size of the 3D reconstruction has a negligible impact on the value of $D_{\rm TV, min}$, while reducing the standard deviation of the Gaussian smoothing kernel results in a higher value of $D_{\rm TV,min}$. This suggests that the asymptotic behavior of $D_{\rm TV}$ does not arise from the limited resolution of the 3D reconstruction but is rather an intrinsic constraint of galaxy-based surveys, likely reflecting the fact that galaxies are biased tracers of the underlying matter distribution and will not perfectly trace the LSS \citep[e.g.,][]{Im2024}.}

Based on these findings, we conclude that the depth of the ODIN survey is well optimized for tracing the LSS, achieving $\Sigma_{\rm LAE}\sim 0.1~{\rm arcmin}^{-2}$. Doubling the source density of ODIN, which would require an increase in the $N501$ depth of $\sim$0.4 mags \citep[based on the luminosity function of][]{Gronwall2007}, would improve the $D_{\rm TV}$ value only by $\lesssim$ 10\%. 



\subsection{Descendant Mass Estimates} \label{subsec:descendant_mass}

\begin{figure*}
    \centering
    \includegraphics[width=0.8\linewidth]{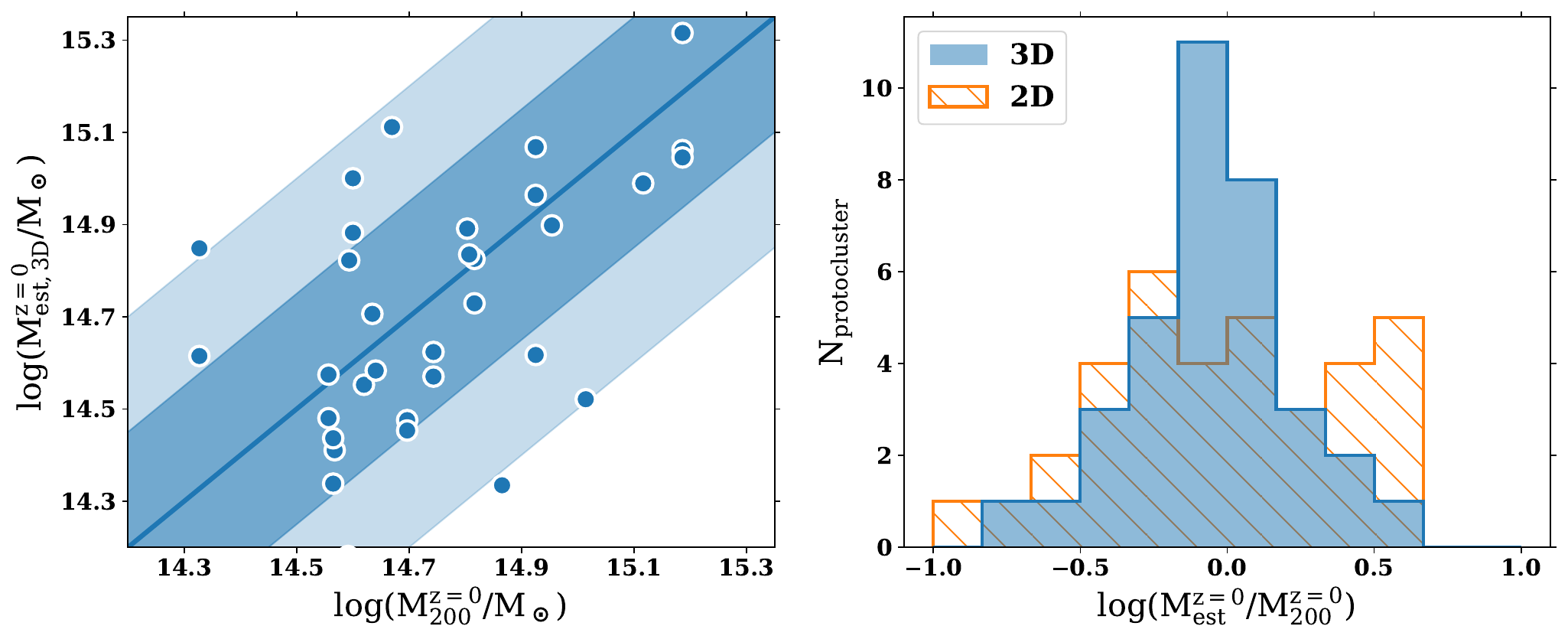}
    \caption{\emph{Left:} Descendant mass at $z=0$ of the 37 structures considered in our analysis, selected from the 30 most massive protoclusters in TNG300 viewed along three sightlines ($x$-axis), compared with the descendant mass estimated from the corresponding 3D reconstruction ($y$-axis). The descendant mass estimates from the reconstruction are corrected for the overestimation of galaxy density due to peculiar motion (see text for details). The solid line shows a one-to-one relation of estimated and true mass, while the darker(lighter) shaded region indicates a range of 0.25(0.5) dex around the one-to-one relation. \emph{Right:} Ratio of estimated to true descendant mass for both the 2D and 3D density maps (shown in log scale). The spread around a ratio of 1 (i.e., the case when the estimated and true descendant masses match perfectly) for the 3D case is much smaller than that for the 2D case.}
    \label{fig:3d_mass}
\end{figure*}

How much improvement does spectroscopic information provide in estimating accurate descendant masses? In \citet{Ramakrishnan2024}, we used the 30 most massive cosmic structures in the TNG300 simulation to calibrate 2D-based mass estimates, $M_{est,{\rm 2D}}^{z=0}$ against true descendant masses. The latter is taken to be $M_{200}^{z=0}$, the mass enclosed within a radius where the matter density is 200 times the critical density. We used Equation~\ref{eq:descendant_mass} with the assumption that protoclusters have roughly isotropic dimensions, i.e., $V_{\rm PC}\approx A_{\rm PC}^{1.5}$, where $A_{\rm PC}$ is the projected surface area of a protocluster. This assumption, necessitated by the lack of better information, is clearly at odds with observational evidence. We found that the resulting large scatter of $\approx$0.5~dex in the mass estimates is primarily driven by the uncertainty in the line-of-sight extent, which is needed to estimate the full 3D volume.


We compare the 2D- and 3D-based estimates of descendant masses for TNG structures under observational conditions matched to COSMOS-z3.1-A, specifically adopting the same LAE surface density ($\bar{\Sigma}_{\rm LAE}=0.1~{\rm arcmin}^{-2}$) and a spectroscopic fraction of $f_{\rm spec}=0.4$. Each TNG structure is `observed' along the X-, Y-, and Z-directions, producing three independent 2D surface density maps and corresponding 3D reconstructions. From these, we derive descendant mass estimates, $M^{z=0}_{est,2D}$ and $M^{z=0}_{est,3D}$, following the same methodology applied to the real data. For the 3D case, the protocluster volume $V_{\rm PC}$ is defined as the volume enclosing all voxels with density values in the top 3rd percentile. This threshold is adopted as it most closely matches the `true' protocluster volume, estimated as described in \citet{Ramakrishnan2024}. However, we find that the descendant mass found with Equation \ref{eq:descendant_mass} is slightly overestimated (by a factor of $\sim$ 2). We attribute this to peculiar motion, which for a collapsing structure has the effect of increasing the observed galaxy density \citep[the Kaiser effect,][]{Kaiser1984}. We apply the appropriate correction to all of our descendant mass estimates. If the resulting $V_{\rm PC}$ is smaller than 500~cMpc$^3$, we consider the structure to be undetected in 3D and exclude it from further analysis. Similarly, if the surface area of a structure does not exceed $A_{\rm PC}=40$~cMpc$^2$, it is not detected as a protocluster in 2D and we discard it. The final sample with robust mass estimates in both 2D and 3D consists of 37 objects.

The left panel of Figure~\ref{fig:3d_mass} compares the 3D-based descendant mass estimates to the true cluster masses at the present day, $M^{z=0}_{200}$. The resulting scatter is 0.25~dex, as indicated by the dark blue shaded region. The scatter varies somewhat depending on the specific choice of mock spec-z galaxies, but remains within the range 0.2--0.3~dex. This represents a significant improvement over the 0.5~dex scatter obtained from the 2D-based estimates. The right panel of Figure~\ref{fig:3d_mass} shows the 2D (orange) vs 3D (blue) comparison.
The intrinsic limit on the accuracy of descendant masses estimated based on the galaxy overdensity is 0.15 dex, which we derived in \citet{Ramakrishnan2024} by estimating $V_{\rm PC}$ using the true 3D positions of \emph{all} TNG galaxies with $M_\ast \geq 10^7 M_\odot$ that will merge into the cluster by $z=0$ (see their Figure~8). These results demonstrate that descendant masses can be estimated with high precision from our 3D reconstruction, lending strong confidence to our mass estimates for COSMOS-z3.1-A and COSMOS-z3.1-C.

We also examine how the accuracy of the mass estimates varies with spectroscopic sampling fraction, $f_{\rm spec}$. Interestingly, the scatter in the 3D-based descendant mass remains relatively stable at 0.2--0.3~dex, even when $f_{\rm spec}$ is as low as 20\%, indicating that the estimated volume remains robust under sparse sampling. However, this trend holds primarily for the most massive protoclusters that appear as prominent overdensities in 3D. In contrast, lower-mass systems are more frequently missed at lower spectroscopic completeness: 15\% (40\%) of the 90 clusters -- comprising 30 systems observed along three sightlines -- are undetected at $f_{\rm spec}=1.0$ (0.2). This incompleteness is reflected in the left panel of Figure~\ref{fig:3d_mass}, where few structures below $\log (M_{200}^{z=0}/M_\odot) \approx 14.6$ are represented even though about half of the 30 most massive structures are below the value \citep[see Table~1 in][]{andrews2024}.

In summary, our findings show that candidate structures identified as strong overdensities in 2D -- such as COSMOS-z3.1-A and C -- are highly likely to be confirmed in 3D, enabling robust reconstruction of their large-scale structure and accurate estimates of their descendant masses, even with relatively sparse spectroscopic coverage. As expected, the likelihood of successful detection and characterization improves for more prominent structures with higher overdensity and/or larger angular extent.


\section{Summary} 


In this work, we present extensive spectroscopic follow-up of COSMOS-z3.1-A and COSMOS-z3.1-C, the two most prominent LAE overdensities identified by the ODIN survey in the extended COSMOS field at $z=3.1$ (see Figure~\ref{fig:spec_sources}). We utilize these data to characterize their 3D spatial structures on the scale of $\approx$50 cMpc in considerable detail. We evaluate our ability to recover the large-scale structure with the available data using the TNG300 suite of cosmological hydrodynamical simulations, and further provide predictions to guide future spectroscopic follow-up of massive structures. Our key findings are summarized below:

\begin{enumerate}

\item We have developed and rigorously validated a 3D reconstruction method that utilizes densely sampled spectroscopic data in highly overdense regions to probabilistically assign three-dimensional positions to {\it all} galaxies in a protocluster (Section~\ref{subsec:method}, Figure~\ref{fig:3d_prior}). We demonstrate that this method yields a clear, quantifiable improvement in the recovery of the 3D matter distribution over reconstructions based solely on spectroscopically confirmed sources (Sections~\ref{subsec:validation} and \ref{subsec:confirmed_fraction}).

\item The 3D density distributions of COSMOS-z3.1-A and COSMOS-z3.1-C (Figures~\ref{fig:cosmos_n501_s1_3d} and \ref{fig:cosmos_n501_s2_3d}) illustrate how different a structure may appear depending on sightlines. Both structures show preferential directions along which they extend and thus appear more elongated. 
{Long tails extending from highly overdense regions are detected, which may indicate cosmic filaments feeding into a parent halo, as often seen in dark matter simulations.} Both structures consist of multiple interconnected but distinct overdensity peaks (Figures~\ref{fig:cosmos_z3.1_a} and \ref{fig:cosmos_z3.1_c}). Each peak is expected to evolve into a virialized cluster-sized halo, and their estimated descendant masses (Table~\ref{tab:density_peaks}) suggest that both COSMOS-z3.1-A and COSMOS-z3.1-C are the progenitors of ultramassive structures with total mass well above~$10^{15} M_\odot$.

\item We definitively confirm that COSMOS-z3.1-A is a proto-supercluster, akin to {\it Hyperion}  -- the only previously known structure of its kind at $z\sim 2.4$ -- though observed here at an earlier cosmic epoch, $z = 3.1$. Like {\it Hyperion}, COSMOS-z3.1-A contains several distinct overdensity peaks (seven in the former and ten in the latter); both span comparable comoving volumes of $\sim$100,000~cMpc$^3$ {and have total enclosed masses $\log(M/M_\odot) \gtrsim 15.5$}. However, they differ significantly in their orientation relative to the observer. {\it Hyperion} appears most extended along the redshift (line-of-sight) axis, whereas COSMOS-z3.1-A is stretched predominantly in the transverse direction. These differences 
emphasize the necessity of combining spectroscopic and photometric approaches to achieve a complete and unbiased census of the largest structures in the Universe.


\item Our 3D maps offer the strongest evidence to date that Ly$\alpha$ blobs preferentially reside either in the outskirts of the highest density peaks or along filaments. While the surface density of LABs in COSMOS-z3.1-A and COSMOS-z3.1-C is $\sim$4 times higher than the average, none of the confirmed LABs are found to reside within the most overdense regions of the structures. The median separation of the LABs from the center of the nearest density peak is $\sim$17~cMpc, placing them well outside the core, given that the characteristic size of the density peaks is 10--20~cMpc. 

\item To guide future spectroscopic and photometric studies of the large-scale structure, we use the TNG300 simulation to assess the dependence of the quality of the 3D density map on the fraction of sources with spectroscopic redshifts ($f_{\rm spec}$) and the mean LAE surface density of the survey ($\bar{\Sigma}_{\rm LAE}$). For the $\bar{\Sigma}_{\rm LAE}$ achieved by ODIN, $f_{\rm spec}$ values as low as 0.2 yield reliable 3D reconstructions that significantly and consistently outperform those based on spectroscopic sources alone (Figure~\ref{fig:conf_frac}). 
Increasing $\bar{\Sigma}_{\rm LAE}$ beyond the ODIN value -- e.g., by doubling exposure time -- provides minimal gain in the LSS recovery if $f_{\rm spec}$ remains fixed. 
With both $\bar{\Sigma}_{\rm LAE}$ and $f_{\rm spec}$ at ODIN levels, descendant halo masses can be estimated with an accuracy of 0.2--0.3~dex, a notable improvement over the 0.5~dex scatter from purely 2D-based approaches (Figure~\ref{fig:3d_mass}).


\end{enumerate}

With its wide area and depth, ODIN is well-positioned to uncover more of these massive cosmic structures at high redshift, particularly at the high-mass end, $\log (M^{z=0}_{200}/M_\odot)\gtrsim 14.6$. 
{The 3D reconstruction methodology presented in this work will enable us to determine the three-dimensional extent and descendant masses of these structures, separate their cores from the outskirts, and identify the galaxies that inhabit these environments.}


\section*{Data Availability}

All data shown in figures are available on Zenodo (doi: \href{https://zenodo.org/records/20756081}{https://doi.org/10.5281/zenodo.20756081}). Interactive 3D visualizations are available on GitHub (\href{https://ortiz140.github.io/odin/}{https://ortiz140.github.io/odin/}).

\begin{acknowledgments}
The authors acknowledge financial support from the U.S. National Science Foundation (NSF) under grant Nos. AST-2206705, AST-2408359, and from the Ross-Lynn Purdue Research Foundations. 
This work is based on observations at Cerro Tololo Inter-American Observatory, NSF’s NOIRLab (Prop. ID 2020B-0201; PI: K.-S. Lee), which is managed by the Association of Universities for Research in Astronomy under a cooperative agreement with the National Science Foundation.
H.S.H.\ acknowledges the support of Samsung Electronic Co., Ltd.\ (Project Number IO220811-01945-01), the National Research Foundation of Korea (NRF) grant funded by the Korea government (MSIT), NRF-2021R1A2C1094577, and Hyunsong Educational \& Cultural Foundation.
R.C.\ and C.G.\ acknowledge support from the National Science Foundation under grant AST-2408358.
The Institute for Gravitation and the Cosmos is supported by the Eberly College of Science and the Office of the Senior Vice President for Research at the Pennsylvania State University.
M.C.A.\ acknowledges support from the ALMA fund with code 31220021 and from ANID BASAL project FB210003.
H.S.\ acknowledges the support of the National Research Foundation of Korea (NRF) grant funded by the Korean government (MSIT) (No. RS-2024-00455106).
M.C.C.\ acknowledges support from a CONICET Doctoral Fellowship.
{L.G.\ gratefully acknowledges financial support from FONDECYT regular project number 1230591, ANID - MILENIO -  NCN2024\_112,  ANID BASAL project FB210003.} \\
This material is based upon work supported by the U.S. Department of Energy (DOE), Office of Science, Office of High-Energy Physics, under Contract No. DE–AC02–05CH11231, and by the National Energy Research Scientific Computing Center, a DOE Office of Science User Facility under the same contract. Additional support for DESI was provided by the NSF, Division of Astronomical Sciences under Contract No. AST-0950945 to the NSF’s National Optical-Infrared Astronomy Research Laboratory; the Science and Technology Facilities Council of the United Kingdom; the Gordon and Betty Moore Foundation; the Heising-Simons Foundation; the French Alternative Energies and Atomic Energy Commission (CEA); the National Council of Humanities, Science and Technology of Mexico (CONAHCYT); the Ministry of Science, Innovation and Universities of Spain (MICIU/AEI/10.13039/501100011033), and by the DESI Member Institutions: \url{https://www.desi.lbl.gov/collaborating-institutions}. Any opinions, findings, and conclusions or recommendations expressed in this material are those of the author(s) and do not necessarily reflect the views of the U. S. National Science Foundation, the U. S. Department of Energy, or any of the listed funding agencies.
The authors are honored to be permitted to conduct scientific research on I’oligam Du’ag (Kitt Peak), a mountain with particular significance to the Tohono O’odham Nation.
\end{acknowledgments}

\bibliography{myrefs}{}
\bibliographystyle{aasjournal}

\appendix

\section{Priors for 3D reconstruction} \label{appendix:priors}


A key requirement of our probabilistic reconstruction methodology, as discussed in the main text, is that the identified structures are genuine and not artifacts introduced by arbitrary binning or smoothing choices. In this section, we detail our approach to selecting appropriate bin sizes and smoothing scales, and assess the robustness of our reconstruction method.

Initially, we construct 3D redshift priors using spectroscopically confirmed sources, grouped into bins of fixed comoving volume, ensuring each XY slice contains a range of redshift values. The primary consideration in selecting bin sizes is to maintain sufficient source density while minimizing the number of empty bins. Additionally, bin size selection must account for the physical scale of protoclusters. Since typical protoclusters span only 10--15~cMpc in diameter \citep[e.g.,][]{Chiang2013}, choosing excessively large bin sizes would effectively place entire protoclusters into single bins, eliminating any meaningful structural information about their internal distribution. If bins are too small, the resulting distribution becomes sparse and noisy. 

To reduce the noise while preserving real features, we applied Gaussian smoothing to our binned distributions. Our initial approach involved smoothing using a three-dimensional Gaussian kernel with a scale of 2--3~cMpc. 
To investigate if smoothing creates artifacts, we also tested an alternative approach where smoothing was applied only along the redshift direction. This method aimed to minimize the influence of projected sources from different line-of-sight positions while still accounting for observational uncertainties in redshift assignment. 

\begin{figure*}
    \centering
    \includegraphics[width=\linewidth]{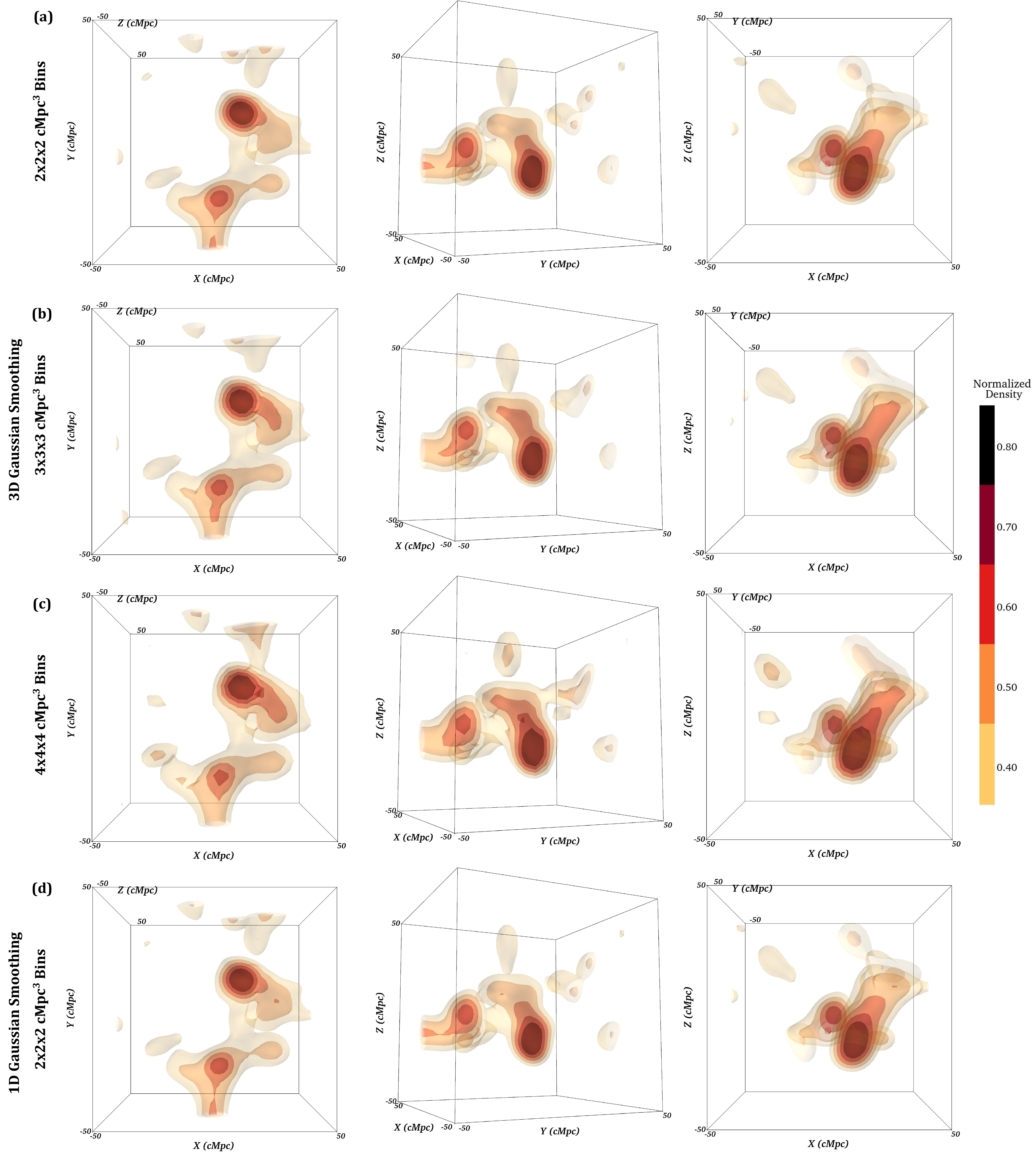}
    \caption{The effect of bin size and smoothing choices on the reconstructed structure of COSMOS-z3.1-C. Each row corresponds to a different reconstruction method and each column shows the reconstruction from a different viewpoint. Rows (a), (b), and (c) use 3D Gaussian smoothing with bin sizes of $(2~{\rm cMpc})^3$, $(3~{\rm cMpc})^3$, and $(4~{\rm cMpc})^3$, respectively. Row (d) applies 1D Gaussian smoothing along the redshift direction while using a $(2~{\rm cMpc})^3$ bin size. While larger bin sizes (row c) result in lower resolution and blockier contours, the overall structures remain consistent. Comparing 3D smoothing (row a) to 1D smoothing (row d), we find that the large-scale structure is preserved, though 1D smoothing introduces slightly greater contour variations.}
    \label{fig:recon_test}
\end{figure*}

In Figure~\ref{fig:recon_test}, we illustrate how COSMOS-z3.1-C appears depending on the bin size and smoothing choices. Rows (a), (b), and (c) show reconstructions using 3D Gaussian smoothing with bin sizes of $(2~{\rm cMpc})^3$, $(3~{\rm cMpc})^3$, and $(4~{\rm cMpc})^3$, respectively. Row (d) uses 1D Gaussian smoothing along the redshift direction while keeping a $(2~{\rm cMpc})^3$ bin size. Across all 3D-smoothed cases (rows a-c), the reconstructed structures remained highly similar, with only minor differences. Larger bin sizes (row c) result in blockier contours due to reduced resolution, but the overall structures persisted. Comparing 3D versus 1D smoothing (rows a and d), we observe that the LSS remained largely unchanged. However, 1D smoothing (row d) exhibits slightly greater variation in density contours, likely due to the lack of the XY-plane smoothing, which would otherwise suppress noise. 

%

Based on these tests, we adopted a bin size of $(2~{\rm cMpc})^3$ with 3D Gaussian smoothing for the final visualizations presented in this paper. This choice balances resolution, noise reduction, and physical justification (that gravitational interactions influence galaxy distributions in all three spatial dimensions). Additionally, keeping both the bin size and smoothing scale well below the typical protocluster size ensures real structures are not erased or oversimplified. The excellent agreement between different bin sizes and smoothing methods provides confidence that our reconstruction reliably represents underlying structures rather than processing artifacts.

\section{Effect of redshift uncertainty} \label{appendix:peculiar_motion}

A major source of uncertainty in positioning galaxies along the line-of-sight direction comes from their peculiar velocities, which result in redshift-space distortion of structures. As a result, the observed redshift of a source will be offset from its cosmological redshift, which in turn will affect the estimate of its position along the sightline. At low redshifts, this leads to the well-known `Finger-of-God' effect \citep{Jackson1972} in observations of galaxy clusters. At high redshift, in the vicinity of overdense but unvirialized protoclusters, the redshift space distortion is instead expected to result in the opposite effect \citep[the Kaiser effect,][]{kaiser87}, wherein the overdensity is enhanced and appears more compact along the line-of-sight due to the infalling motion of the galaxies. 

Another possible source of uncertainty comes from the fact that Ly$\alpha$ emission is generally observed to be redshifted relative to the systemic velocity \citep[e.g.,][]{Erb2014,Muzahid2020}. Moreover, this velocity offset between the Ly$\alpha$ velocity and systemic velocity is not uniform, but is rather a function of the galaxy properties. \citet{Erb2014} find that the velocity offset is negatively correlated with the equivalent width (EW) of the Ly$\alpha$ line, which they explain on the basis that an increase in the column density of neutral gas has both the effect of suppressing the strength of Ly$\alpha$ emission and redshifting the escaping Ly$\alpha$ photons, as a result of increased scattering. \citet{Muzahid2020} meanwhile find a positive correlation of the velocity offset and star formation rate (SFR). These correlations may lead to coherent gradients in the velocity offset from the protocluster outskirts to the core, if there are also coherent trends between the mass and/or SFR of galaxies and their position within the overdense region.

In this section, we seek to understand how significant of an effect both redshift-space distortions and the Ly$\alpha$ velocity offset have on our reconstruction. As before, we consider a $(60~{\rm cMpc})^3$ region centered on the $z~=~3$ progenitor of a massive cluster. We designate a fraction of the mock LAEs in this region as spec-z sources and probabilistically assign the remaining mock photo-z galaxies a redshift. We repeat this redshift assignment process 500 times and average over all 500 realizations to arrive at the final 3D reconstruction. However, for a given set of mock spec-z galaxies, we now consider three redshift values: 1) the `true' cosmological redshift, which is based only on the galaxy's position along the line-of-sight axis, as basis for comparison 2) the `observed' redshift, which is perturbed from the true redshift according to the peculiar velocity of each galaxy along the line-of-sight axis and 3) the `Ly$\alpha$' redshift, which is further redshifted from the observed value as discussed in the following paragraph. By comparing the 3D reconstructions using these three redshift values for the same galaxies, we can assess the effect of these uncertainties in the measured redshift on our 3D reconstruction.

To determine the additional velocity offset used to mimic the behavior of the Ly$\alpha$ line, we use the relationship between the Ly$\alpha$ velocity offset, $v_{\rm offset}$, and SFR determined by \citet{Muzahid2020}: 
\begin{equation}\label{eqn:v_offset}
\log(v_{\rm  offset}/{\rm km~s^{-1}}) = 0.16 \times \log({\rm SFR}/M_\odot~{\rm yr^{-1}}) + 2.26    
\end{equation}
We note that our results remain unchanged both on fixing $v_{\rm offset}$ to the median value obtained for the mock LAEs from Equation \ref{eqn:v_offset}, and upon increasing the second term of Equation \ref{eqn:v_offset} from 2.26 to 2.50 (i.e. upon increasing the median value of $v_{\rm offset}$ for the sample).

\begin{figure}
    \centering
    \includegraphics[width=0.5\linewidth]{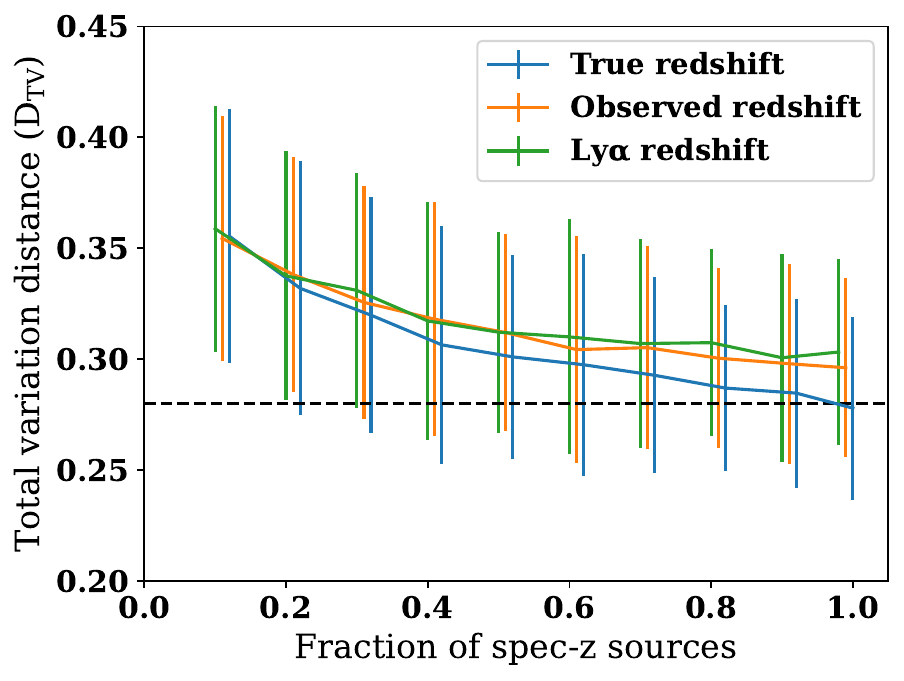}
    \caption{Total variation distance $D_{\rm TV}$ of the 3D reconstruction of the progenitors of the 30 most massive clusters in TNG300 \citep{andrews2024}, relative to the dark matter distribution. The blue line indicates $D_{\rm TV}$ for the 3D reconstruction carried out without incorporating the effect of peculiar motion on the observed redshift. The orange indicates that for the 3D reconstruction which does incorporate this effect. The green line indicates $D_{\rm TV}$ for the case which incorporates both the peculiar motion of the galaxies and the generally observed redshift of the Ly$\alpha$ line away from the systemic velocity. The horizontal dashed line indicated the minimum possible value of $D_{\rm TV}$ for our LAE surface density, i.e. assuming 100\% spectroscopic confirmation and no offsets in the measured redshifts.}
    \label{fig:tvd_pec}
\end{figure}

We quantify the difference between the three cases in Figure \ref{fig:tvd_pec}, where we show the $D_{\rm TV}$ of the 3D reconstructions using true redshifts (blue), observed redshifts (orange) and Ly$\alpha$ redshifts (green) relative to the dark matter distribution. The possible spread of the $D_{\rm TV}$ values, indicated by the errorbars, is estimated by repeating the reconstruction procedure for 90 different TNG300 protoclusters (namely the progenitors of the 30 most massive clusters viewed along 3 different sightlines, see Section \ref{subsec:descendant_mass}). It can be seen that while the peculiar motion of the galaxies does on average increase the $D_{\rm TV}$ value (indicating a larger departure from the dark matter distribution) for all spectroscopic fractions, the magnitude of this effect is smaller than the scatter. Including the Ly$\alpha$ velocity offset slightly increases $D_{\rm TV}$ relative to the case which uses only the peculiar velocity, but this effect is very small. 

For the TNG300 protoclusters we consider, the typical peculiar velocities of the galaxies along the line-of-sight are in the range 200 - 300 km/s, resulting in an offset of 2.5--4~cMpc in the position. The magnitude of the Ly$\alpha$ velocity offset is on average lower than that of the peculiar velocity, $\sim$150--200~km~s$^{-1}$. The uncertainty in the line-of-sight position arising from these effects is much smaller than the typical size of the structures themselves, $\sim$10--20~cMpc, which is likely the reason that the $D_{TV}$ value measured relative to the dark matter distribution is similar in all three cases. Thus, we conclude that neither peculiar velocities nor the systematic Ly$\alpha$ velocity offset poses a significant concern to our 3D reconstruction. 

\end{document}